\documentclass[12pt]{article}
\usepackage[english]{babel}
\usepackage{amsfonts,amsmath,amssymb}
\usepackage{mathtools}
\usepackage{graphics}
\usepackage{tikz}
\usepackage{subfigure}
\usepackage[font=small,labelfont=bf]{caption}
\usepackage{booktabs}
\usepackage{wasysym}
\usepackage{multirow}
\usepackage{enumerate}
\usepackage{epstopdf}
\usepackage{bbold}
\usepackage{bbm}
\usepackage{tabularx}
\usepackage{lipsum}
\usepackage{ragged2e}  
\usepackage{booktabs} 
\usepackage{hyperref}
\usepackage{geometry}
\newcommand{\re}{\operatorname{Re}}

\newcommand{\mpcac}{m_\mathrm{PCAC}}

\newcommand{\csw}{c_\mathrm{SW}}

\newcommand{\tr}{\mathrm{Tr}}

\usetikzlibrary{arrows,positioning}
\tikzstyle{line} = [draw, -]

\title{Simulating twisted mass fermions at physical light, strange and charm quark masses}
\author{Constantia Alexandrou\textsuperscript{a,b}, Simone Bacchio\textsuperscript{a,c}, \\  Panagiotis Charalambous\textsuperscript{b},
Petros Dimopoulos\textsuperscript{d,e}, Jacob Finkenrath\textsuperscript{b},\\  Roberto Frezzotti\textsuperscript{d},
Kyriakos Hadjiyiannakou\textsuperscript{b}, Karl Jansen \textsuperscript{f}, \\ Giannis Koutsou\textsuperscript{b}, 
Bartosz Kostrzewa\textsuperscript{g}, Mariane Mangin-Brinet\textsuperscript{h},\\ Giancarlo Rossi\textsuperscript{d,e},
Silvano Simula\textsuperscript{i}, Carsten Urbach\textsuperscript{g}\\
\small\textsuperscript{a}Department of Physics, University of Cyprus, PO Box 20537, 1678 Nicosia, Cyprus\\
\small\textsuperscript{b}Computation-based Science and Technology Research Center,\\ \small The Cyprus Institute,  20 Konstantinou Kavafi Street, 2121 Nicosia, Cyprus\\
\small\textsuperscript{c}Fakult\"at f\"ur Mathematik und Naturwissenschaften,\\  \small Bergische Universit\"at Wuppertal,  Gau{\ss}str.~20, 42119 Wuppertal \\
\small\textsuperscript{d}Dip. di Fisica, Universit{\`a} and INFN di Roma Tor Vergata, 00133 Roma, Italy \\
\small\textsuperscript{e}Centro Fermi - Museo Storico della Fisica e Centro \\  \small  Studi e Ricerche “Enrico Fermi’, Rome, Italy \\
\small\textsuperscript{f}NIC, DESY, Zeuthen, Platanenallee 6, 15738 Zeuthen, Germany \\
\small\textsuperscript{g}HISKP (Theory), Rheinische Friedrich-Wilhelms-Universit{\"a}t Bonn,\\ \small Nu{\ss}allee 14-16,  53115 Bonn, Germany  \\
\small\textsuperscript{h}Theory Group, Lab.~de Physique Subatomique et de Cosmologie, \\ \small 38026 Grenoble, France \\
\small\textsuperscript{i}Istituto  Nazionale  di  Fisica  Nucleare,  Sezione  di  Roma  Tre, \\ \small Via  della  Vasca  Navale  84, I-00146  Rome,  Italy \\ \\
\includegraphics[width = 25mm]{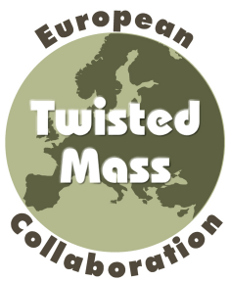}
\vspace{-0.6cm}
}

  \date{}

\begin{document}

\maketitle
\begin{abstract}
We present the QCD simulation of the first  gauge ensemble of two degenerate light quarks, a strange  and a charm quark  
with all quark masses tuned to their physical values within the twisted mass fermion formulation.
Results for the pseudoscalar masses and decay constants 
confirm that the produced ensemble is indeed at the physical parameters of the theory.
This conclusion is corroborated by a complementary analysis in the baryon 
sector. We examine cutoff and isospin breaking  effects and demonstrate that they are suppressed 
through the presence of a clover term in the action.
\end{abstract}

\newpage
\section{Introduction\label{sec:introduction}}

Simulations of Quantum Chromodynamics    
directly with  physical quark masses, large enough volume and small enough lattice spacing have become 
feasible due to significant algorithmic improvements and
availability of substantial computational resources. 
In fact, state-of-the-art simulations
using different discretization schemes are currently being carried out 
worldwide.

Within the twisted mass formulation~\cite{Frezzotti:2000nk,Frezzotti:2003ni,Frezzotti:2004wz}, 
the European Twisted Mass Collaboration (ETMC) has carried out 
simulations directly at the physical value
of the pion mass~\cite{Abdel-Rehim:2015owa,Abdel-Rehim:2015pwa} with $N_f=2$ mass-degenerate up and down
quarks at a lattice spacing of $a=0.0913(2){\rm fm}$. This is a remarkable result, since 
explicit isospin breaking effects associated with  twisted mass fermions 
can make physical point simulations at too coarse values of the
lattice spacing  very difficult. 
Being able to reach the physical pion mass for the case of $N_f=2$ flavours
was therefore of great importance and many physical quantities
have already been computed on the so generated gluon field configurations.
Examples are meson properties \cite{Abdel-Rehim:2015pwa,Abdel-Rehim:2015owa,Liu:2016cba,Liu:2017nzk,Helmes:2017ccf,Alexandrou:2017blh}, 
the structure of hadrons 
\cite{Abdel-Rehim:2015owa,Abdel-Rehim:2016won,Alexandrou:2017hac,Alexandrou:2017ypw,Alexandrou:2017oeh,Alexandrou:2017dzj}  
and the anomalous magnetic moment of the 
muon \cite{Abdel-Rehim:2015pwa}.  

The success of these $N_f=2$ flavour simulations
strongly suggests to extend the calculation by adding the
strange and charm quarks as dynamical degrees of freedom,
a situation we will refer to as  $N_f=2+1+1$ simulation. Adding a quark doublet is
a natural step for twisted mass fermions. 
However, it is known that
the presence of a heavy quark doublet in the sea gives rise to larger discretization effects than
having only the light up and down quarks.

This paper reports on our successful, but demanding tuning effort
to reach a physical situation with the first two quark generations tuned to their physical values
in the twisted mass representation. We will present
first results for low-lying meson masses and decay constants as well as baryon masses. 
In addition, we describe a comprehensive determinations of the 
lattice spacing from the meson and baryon sectors as well as from gradient 
flow observables. Furthermore, we discuss isospin 
breaking effects of twisted mass fermions in the neutral and charged 
pion and in the $\Delta$ sector.
Demonstrating the
successful generation of an  $N_f=2+1+1$ ensemble of  maximally twisted mass fermions
at physical quark  masses is 
the essential result of this paper that
lays the ground for a future very rich
research program within the twisted mass formulation 
with an eventually  large impact for ongoing and planned experiments.

The  outline of the paper is as follows:
In section~\ref{sec:action}
we introduce the employed twisted mass action and discuss
details of the  parameters used in the Hybrid Monte Carlo simulation.
In  section~\ref{sec:tun} we discuss our tuning procedure to reach physical
light, strange and charm quark masses, which includes tuning for $\mathcal{O}(a)$
improvement and a discussion on the isospin splitting.
In  section~\ref{sec:mesons} we present mesonic
quantities for our ensemble, including a determination of the lattice spacing
via the pion decay constant and heavy quark observables.
In section \ref{sec:bar} we discuss nucleon properties  including
the determination of the lattice spacing via the nucleon mass. In addition, we discuss possible  isospin splitting
in the $\Delta$-baryon sector. In section~\ref{sec:olatsp} 
we summarize the different determination of lattice spacing
via gluonic, mesonic and baryonic observables and conclude.

\section{Action}
\label{sec:action}

We employ the twisted mass fermion formulation, within which observables 
are automatically $\mathcal{O}(a)$ improved when working at 
maximal twist~\cite{Frezzotti:2003ni,Frezzotti:2003xj}. This formulation
has proven very advantageous: 
It allows one to perform safe, infrared regulated
simulations and simplified renormalization in some cases. There is no need for
improvement on the operator level due to automatic $\mathcal{O}(a)$-improvement and cut-off
effects turn out to be relatively small except for the special case of the neutral (unitary) pion mass.

\begin{table}[h]
  \centering
  \begin{tabular}{ccccccc }
    \hline \hline \\[-2.0ex]
  V               & $\beta$        &    $\mu_\ell$   & $\mu_\sigma$  & $\mu_\delta$              &  $\kappa$     &   $\csw$             \\[0.8ex]
    \hline \\[-1.8ex]
 $128\times64^3$ &   1.778         &  0.00072     &   0.1246864      & 0.1315052                           &  $0.1394265$   &   $1.69$      \\[0.8ex]
    \hline \hline 
  \end{tabular}
  \caption{The table shows the parameters which are used in the 
simulations of the target ensemble labeled by cB211.072.64 with pion masses close 
  to the physical point.}
  \label{tab:parametersL64}
\end{table}

The action of twisted mass fermions is given by
\begin{equation}
    S = S_g + S_{tm}^\ell  + S_{tm}^h ~ ,
    \label{eq:totalaction}
\end{equation}
where we choose the Iwasaki improved gauge action for $S_g$ \cite{Iwasaki:1985we} 
which reads
\begin{equation}
\label{eq:iwasaki}
S_g = \frac{\beta}{3}\sum_x\left(  b_0\sum_{\substack{
    \mu,\nu=1\\1\leq\mu<\nu}}^4\{1-\re\tr(U^{1\times1}_{x,\mu,\nu})\}\Bigr.
\Bigl.\ +\
b_1\sum_{\substack{\mu,\nu=1\\\mu\neq\nu}}^4\{1
-\re\tr(U^{1\times2}_{x,\mu,\nu})\}\right)\,  ,
\end{equation}
with the bare inverse gauge coupling $\beta=6/g_0^2$, $b_1=-0.331$ and
$b_0=1-8b_1$.
 In the 
case of the light up and down quark doublet, the action takes the form
\begin{equation}
  \label{eq:sflight}
  S_{tm}^\ell = \sum_x \bar{\chi}_\ell(x)\left[ D_W(U) + \frac{i}{4} \csw \sigma^{\mu\nu}
    \mathcal{F}^{\mu\nu}(U) + m_\ell + i \mu_\ell \tau^3 \gamma^5 
  \right] \chi_\ell(x)\,.
\end{equation}
Here, $\chi_\ell=(u,d)^t$ represents the light quark doublet, $\mu_\ell$ is the twisted and 
$m_\ell$ the (untwisted) Wilson quark mass. The Pauli matrix $\tau_3$ acts in flavour 
space 
and $D_W$ is the massless Wilson-Dirac operator. Note that the Wilson quark mass
$m_\ell$ and the clover term $\frac{i}{4} \csw \sigma^{\mu\nu}\mathcal{F}^{\mu\nu}(U)$ 
--with Sheikoleslami-Wohlert improvement coefficient  $c_\mathrm{SW}$ \cite{Sheikholeslami:1985ij}--
are trivial in flavour space. 

For the heavy quark action, with mass non-degenerate strange (s) and charm (c)
quarks, we construct a quark doublet $\chi_h=(s,c)^t$ for which 
the action reads~\cite{Frezzotti:2003xj}
\begin{equation}
  \label{eq:sfheavy}
  S_{tm}^h = \sum_x \bar{\chi}_h(x)\left[ D_W(U) + \frac{i}{4} \csw \sigma^{\mu\nu}
    \mathcal{F}^{\mu\nu}(U) + m_h - \mu_\delta \tau_1 + i \mu_\sigma \tau^3 \gamma^5
  \right] \chi_h(x)\,.
\end{equation}
The important addition compare to eq.~(\ref{eq:sflight}) is the term 
$\mu_\delta \tau_1$ with $\tau_1$ again acting in flavour space. 
The Wilson quark masses in eqs.~(\ref{eq:sflight},\ref{eq:sfheavy}) 
are related to the hopping parameter $\kappa$ as
$m = 1/2\kappa - 4$.
By tuning the light Wilson bare quark mass $m_\ell$ to its critical value $m_{crit}$
the maximally twisted fermion action is obtained for which all physical 
observables are automatically $\mathcal{O}(a)$-improved \cite{Frezzotti:2003ni,Frezzotti:2003xj}.
Setting $m_h = m_\ell = m_{crit}$ this property takes over to the heavy quark mass action
such that only one bare mass parameter has to be tuned to its critical value which
is a great simplification for practical simulations.

However quadratic lattice artefacts can be sizable but by introducing a clover term 
they can be suppressed, e.g.~in case of the neutral pion mass
as shown in \cite{Becirevic:2006ii,Dimopoulos:2009es,Abdel-Rehim:2015pwa,Finkenrath:2017}.
Here, the clover parameter is set by using an estimate from 1--loop \cite{Aoki:1998qd}
tadpole boosted perturbation theory given by
\begin{equation}
c_{SW} \cong 1+ 0.113(3) \frac{g_0^2}{P} 
\end{equation}
with $P$ the plaquette expectation value. For our target parameter set, shown in tab.~\ref{tab:parametersL64},
the plaquette expectation value is given by $P=0.554301(6)$, which is consistent with setting $c_{SW} = 1.69$.

\subsection{Algorithm}
\label{subsec:alg}

For the generation of the  gauge field configurations we use 
as a basis the Hybrid Monte Carlo (HMC) algorithm~\cite{Gottlieb:1987mq,Duane:1987de}
as described in Ref.~\cite{Urbach:2005ji,Jansen:2009xp}. 
For the light quark
sector Hasenbusch mass preconditioning~\cite{Hasenbusch:2001ne,Hasenbusch:2002ai} is applied.
In particular, we employ four determinant ratios with  
mass shifts $\rho = \{0.0;\, 0.0003;\, 0.0012;\, 0.01;\, 0.1\}$.
The heavy quark determinant is treated by a rational approximation \cite{Clark:2006fx,Luscher:2010ae}  
with 10 terms tuned such that the (eigenvalue) interval 
$[0.000065 , \, 4.7]$ is covered. 
For the molecular dynamics integration we
use a nested second order minimal 
norm integrator. This results
in 12 integration steps for the smallest mass term in the light
and heavy quark sector and 192 steps for the gluonic sector~\cite{Finkenrath:2017}.
We use the software package tmLQCD~\cite{Jansen:2009xp} which incorporates the multi-grid algorithm
DDalphaAMG for the inversion of the Dirac matrix~\cite{Frommer:2013fsa}. The force
calculation in the light quark sector is accelerated by a 3-level multi-grid 
approach optimized for twisted mass fermions~\cite{Alexandrou:2016izb}. 
Moreover, we extended the DDalphaAMG method for the mass non-degenerate twisted mass
operator. The multi-grid solver used in the rational approximation~\cite{Bacchio:2017,Bacchio:2017a}
is particularly helpful for 
the lowest terms of the  rational approximation, as well as for the rational 
approximation corrections in the acceptance steps, where it 
yields a speed up of two over the standard multi-mass shifted
conjugate gradient(MMS-CG) solver.
We checked the size of reversibility violation of this setup
yielding a standard deviation $<0.01$ for $\delta \Delta H$ 
and $| 1 -\langle \Delta H \rangle| < 0.02$ 
fulfilling the criteria discussed in \cite{Urbach:2017ate}.
Here, $\delta \Delta H$ is the difference of the Hamiltonian at integration time $t=0$
and the Hamiltonian of the reversed integrated field variables after one trajectory is performed.

\begin{center}
\begin{table}
\begin{center}
\begin{tabular}{lllllll}
\hline \hline 
Ensemble & L & $a \mu_\ell$ & $\kappa$ & $N_{th}$ & $a \mu_\sigma$ & $a \mu_\delta$\\ \hline 
Th1.350.24.k1 & 24 & 0.0035 & 0.1394 & 755 & 0.1162 & 0.1223\\
Th1.350.24.k2 & 24 & 0.0035 & 0.13942 & 350 & 0.1162 & 0.1223\\
Th1.350.24.k3 & 24 & 0.0035 & 0.13945 & 351 & 0.1162 & 0.1223\\
Th1.350.24.k4 & 24 & 0.0035 & 0.13950 & 267 & 0.1162 & 0.1223\\
Th1.350.32.k1 & 32 & 0.0035 & 0.13940 & 88 & 0.1162 & 0.1223\\
Th1.200.32.k2 & 32 & 0.002 & 0.13942 & 430 & 0.1162 & 0.1223\\
Th2.200.32.k1 & 32 & 0.002 & 0.13940 & 178 & 0.1246864 & 0.1315052\\
Th2.200.32.k2 & 32 & 0.002 & 0.13942 & 439 & 0.1246864 & 0.1315052\\
Th2.200.32.k3 & 32 & 0.002 & 0.13944 & 392 & 0.1246864 & 0.1315052\\
Th2.125.32.k1 & 32 & 0.00125 & 0.139424 & 815 & 0.1246864 & 0.1315052\\
cB211.072.64.r1 & 64 & 0.00072 & 0.1394265 & 1647 & 0.1246864 & 0.1315052\\
cB211.072.64.r2 & 64 & 0.00072 & 0.1394265 & 1520 & 0.1246864 & 0.1315052 \\
\hline \hline 
\end{tabular}
\end{center}
\caption{ Summary of the parameters of the ensembles used for the tuning and final runs: 
L is the lattice spatial size with the time direction taken to be $2L$,
$a \mu_\ell$ is the twisted mass parameter of the mass degenerate light quarks,
$\kappa$ is the hopping parameter (common to all flavours), $N_{th}$ are the number of thermalized trajectories in 
molecular dynamics units (MDU),
$a \mu_\sigma$ and $a \mu_\delta$ are the bare twisted mass 
parameter of the mass non-degenerate fermion action used for the heavy quark sector.
The ensembles cB211.072.64.r1 and cB211.072.64.r2 represent the targeted large volume runs at the 
physical point. 
}
\label{tab:over}
\end{table}
\end{center}

\section{Quark Mass Tuning}
\label{sec:tun}

\subsection{Tuning of the light quark sector}
\label{sse:tlq}

As shown in Ref.~\cite{Frezzotti:2005gi,Jansen:2005kk} a most suitable 
and theoretically sound 
condition for the desired automatic 
$\mathcal{O}(a)$ improvement for twisted mass fermions 
is achieved by demanding a vanishing of the partially conserved axial current (PCAC) quark mass
\begin{equation}
  \label{eq:mpcac}
  \mpcac =
  \frac{\sum_\mathbf{x}\langle\partial_0 A^a_0(\mathbf{x},t)P^a(0)\rangle}  {2\sum_\mathbf{x}\langle P^a(\mathbf{x},t)P^a(0)\rangle} \,
  ,\qquad\quad a=1,2\, 
\end{equation} 
with $A_\mu^a$ the axial vector current and $P^a$ the 
pseudoscalar current. In the twisted basis and for light, mass
degenerate quarks, the axial and pseudoscalar currents can be calculated via
\[
A_\mu^+(x) = \bar\chi_\ell(x)\gamma_\mu\gamma_5\frac{\tau^+}{2}\chi_\ell(x)\,
,\qquad\qquad P^+(x) = \bar\chi_\ell(x)\gamma_5\frac{\tau^+}{2}\chi_\ell(x)\, .
\]
using $\tau^+ = (\tau_1 + i\tau_2)/2$ where $\tau_i$ are the Pauli matrices.
The tuning procedure to maximal twist 
requires a value of the hopping parameter $\kappa=\kappa_{\rm crit}$  where 
$\mpcac(\kappa_{\rm crit}) = 0$.
Note that the corresponding
definition of the critical mass $am_{crit} = 1/(2\kappa_{crit}) -4$ is a function of $a\mu_\ell$, $a\mu_\sigma$, $a\mu_\delta$.
Thus, even if the $1/a$ divergence in $m_{crit}$ is independent from $\mu_\ell$, $\mu_\sigma$ and $\mu_\delta$,
determining $am_{crit}$ at the $\mu_\sigma$ and $\mu_\delta$ values of interest is 
important in order to keep lattice artifacts small
which are introduced by the heavy quark doublet \cite{Baron:2010th,Carrasco:2014cwa}.  
Instead the dependence of $am_{crit}$ on $\mu_\ell$ 
reflects much milder discretization errors.
In practice, we allow for some tolerance to this strict condition and 
following Ref.~\cite{Boucaud:2008xu} we impose that
\begin{equation}
 \frac{Z_A \mpcac}{\mu_\ell} < 0.1 
 \label{eq:condPCAC}
\end{equation}
within errors. In eq.~(\ref{eq:condPCAC}) $Z_A$ is the renormalization 
constant of the axial current.
Fulfilling the condition eq.~(\ref{eq:condPCAC})
is numerically consistent with $\mathcal{O}(a)$-improvement of physical observables, 
where it entails only an error of order $\mathcal{O}((Z_A \cdot \mpcac / \mu_\ell)^2)$.
Hence for $< 0.1$ follows for the targeted lattice spacing 
the error is comparable to other $\mathcal{O}([a\Lambda_{QCD}]^2)$ discretization errors.
This allows an  $\mathcal{O}(a^2)$ scaling of physical observables towards the continuum limit.

In order to tune to $\kappa_{\rm crit}$, 
we have generated several ensembles with fixed volumes of 
size $24^3\cdot 48$ and $32^3\cdot 64$, as listed in Table~\ref{tab:over}. For a  
fixed twisted mass parameter of the up and down doublet, we scan over several values of the
hopping parameter $\kappa$, see Table~\ref{tab:over}. 
After fixing  $\kappa_{\rm crit}$ in this manner we proceed 
by tuning the light and heavy twisted mass parameters to realize 
physical pion, kaon and D-meson masses and decay constants. This procedure, 
which is described in more detail below will provide the 
input parameters for the target large volume simulations,
denoted as the ensembles cB211.072.64.r1 and cB211.072.64.r2 in 
Table~\ref{tab:over}.

Initially, we had attempted to  start our $N_f=2+1+1$ simulations at a smaller
value of $\beta=1.726$ that would correspond to the lattice spacing of our $N_f=2$ ensemble with
 $a\sim 0.095 \; \textrm{fm}$ \cite{Abdel-Rehim:2015pwa}. 
However, it turned out that tuning 
to maximal twist for a physical value of the 
pion mass for this $\beta$-value was not feasible. 
Nevertheless, our simulations at $\beta=1.726$ for pion masses in the 
range between $170\;\textrm{MeV}$ and $350\;\textrm{MeV}$
allowed us to develop a tuning strategy to realize the situation of maximal twist 
and also to reach the physical kaon and D-meson masses. 
This tuning strategy was then used at the finer lattice 
spacing as discussed in the present paper.  
The occurrence of instabilities of the simulations at $\beta=1.726$ 
when approaching the physical pion mass
is, in fact, not unexpected. With twisted mass 
fermions, going to sufficiently small values of the light twisted mass 
parameter at a fixed lattice spacing one either enters the Aoki~\cite{Aoki:2004ta} or the 
Sharpe-Singleton~\cite{Sharpe:1998xm} regime, see for an recent overview \cite{Janssen:2015lda}.
For the Sharpe-Singleton case, which is realized in our unquenched simulations,
a sizable $\mathcal{O}(a^2)$ negative shift of the neutral pion mass occurs.

Let us consider the region close to maximal twist,
where $|\omega - \pi/2| \ll 1$ or, equivalently, $m_\ell = m_0 - m_{crit} \ll \mu_\ell$. 
Here the pion mass splitting can be related
the PCAC quark mass by \cite{Aoki:2004ta,Sharpe:2005rq}
\begin{equation}
 a \mpcac \sim Z a m_\ell \frac{m_{\pi}^2}{m_{\pi^{(0)}}^2} + \ldots\; .
 \label{eq:PCACslope}
\end{equation}
where $Z=Z_m Z_P/Z_A$ is a combination of the  untwisted quark mass ($Z_m$), the pseudoscalar 
($Z_P$) and the axial ($Z_A$) renormalization factors and $a m_{0}$ 
denotes the bare quark mass. The charged pion mass is denoted throughout this
paper by $m_{\pi}$, while the neutral pion is given by $m_{\pi^{(0)}}$.
The twisted mass angle $\omega$ can be defined via the gap equation, see \cite{Aoki:2004ta,Sharpe:2005rq}.
From eq.~(\ref{eq:PCACslope}) it is clear that the tuning necessary to satisfied eq.~(\ref{eq:condPCAC})
becomes very hard for a large pion mass difference $m_\pi^2 - {m_{\pi^{(0)}}}^2 \gg 0$. 

In the Sharpe-Singleton scenario a first order phase transition 
is predicted from chiral perturbation theory. In simulations on 
finite lattices this leads to large fluctuations and jumps 
of physical observables~\cite{Farchioni:2004us,Farchioni:2004ma,Farchioni:2004fs,Farchioni:2005ec,Farchioni:2005bh,Farchioni:2005tu}, driving the simulations
to become unstable. This makes it very hard 
to tune successfully to maximal twist. 
In our simulations at $\beta = 1.726$ we observed a strong dependence of the 
PCAC quark mass on the bare mass parameter $m_0$, 
which made it difficult to tune to the critical hopping parameter for a pion mass below $170\; \textrm{MeV}$.
Although at $\beta = 1.726$ we did not investigate in detail which of the lattice ChPT scenario is realized, the fact that at $\beta = 1.778$
we find (see Section \ref{sse:Isosp}) a neutral pion mass $\sim20\%$ smaller than the charged one suggests 
that a Singleton-Sharpe lattice scenario occurs in the scaling region with our chosen action (see Sec.~\ref{sec:action}).

In order to avoid the aforementioned difficulties, we therefore decided to choose 
a finer value of the lattice spacing that would facilitate tuning to critical mass 
at the physical point. 
We found that a value of 
$\beta= 1.778$, corresponding to  $a\approx 0.08 \; \textrm{fm}$, allows us      
to tune to maximal twist successfully.    
In the following, we consider therefore a lattice volume of size 
$64^3\cdot 128$, which 
is sufficiently large to suppress finite size 
effects but at the same time can be simulated with reasonable computational resources, given 
the algorithmic improvements that were discussed in section~\ref{subsec:alg}.

\begin{figure}
\includegraphics[width=0.56\textwidth]{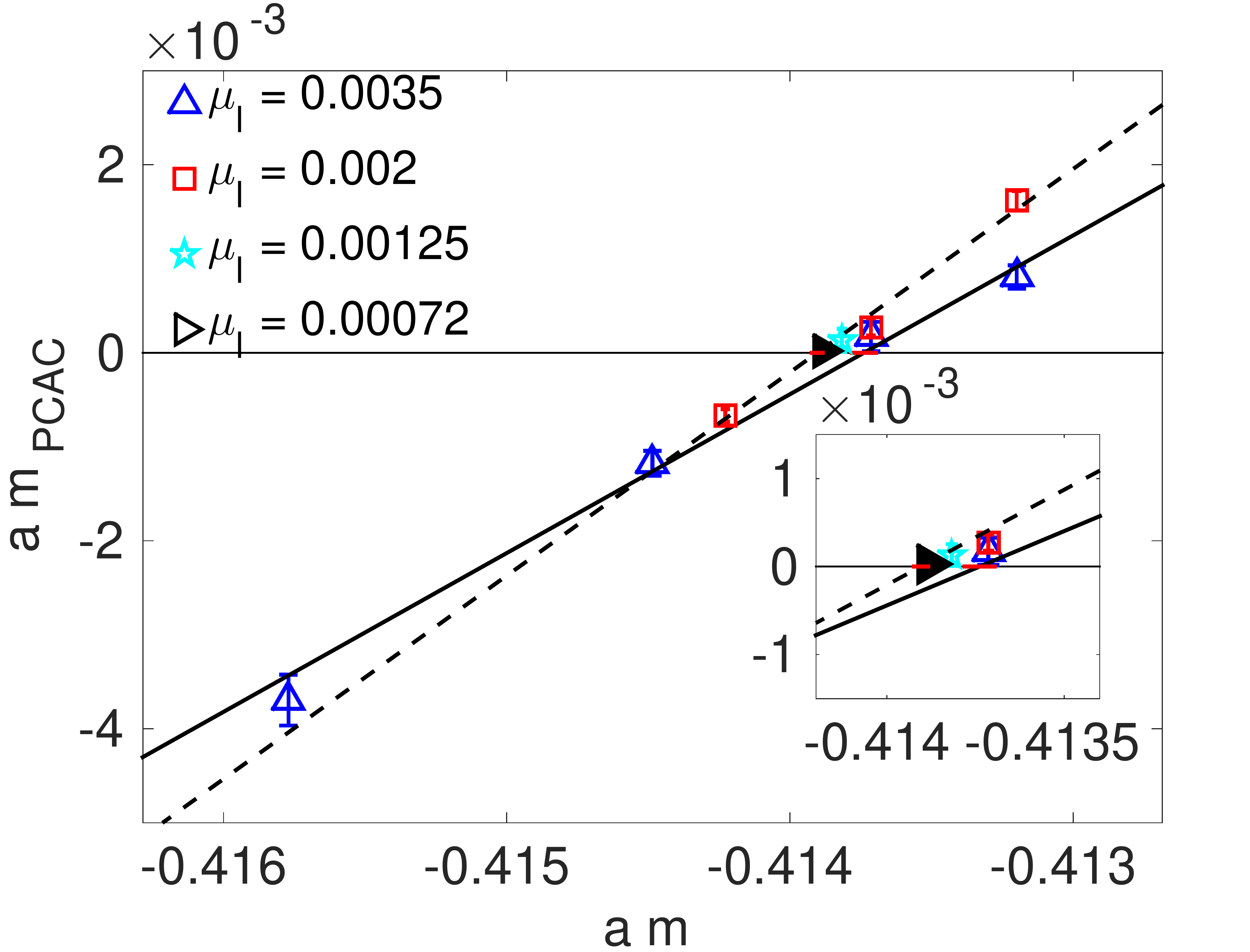}
\includegraphics[width=0.43\textwidth]{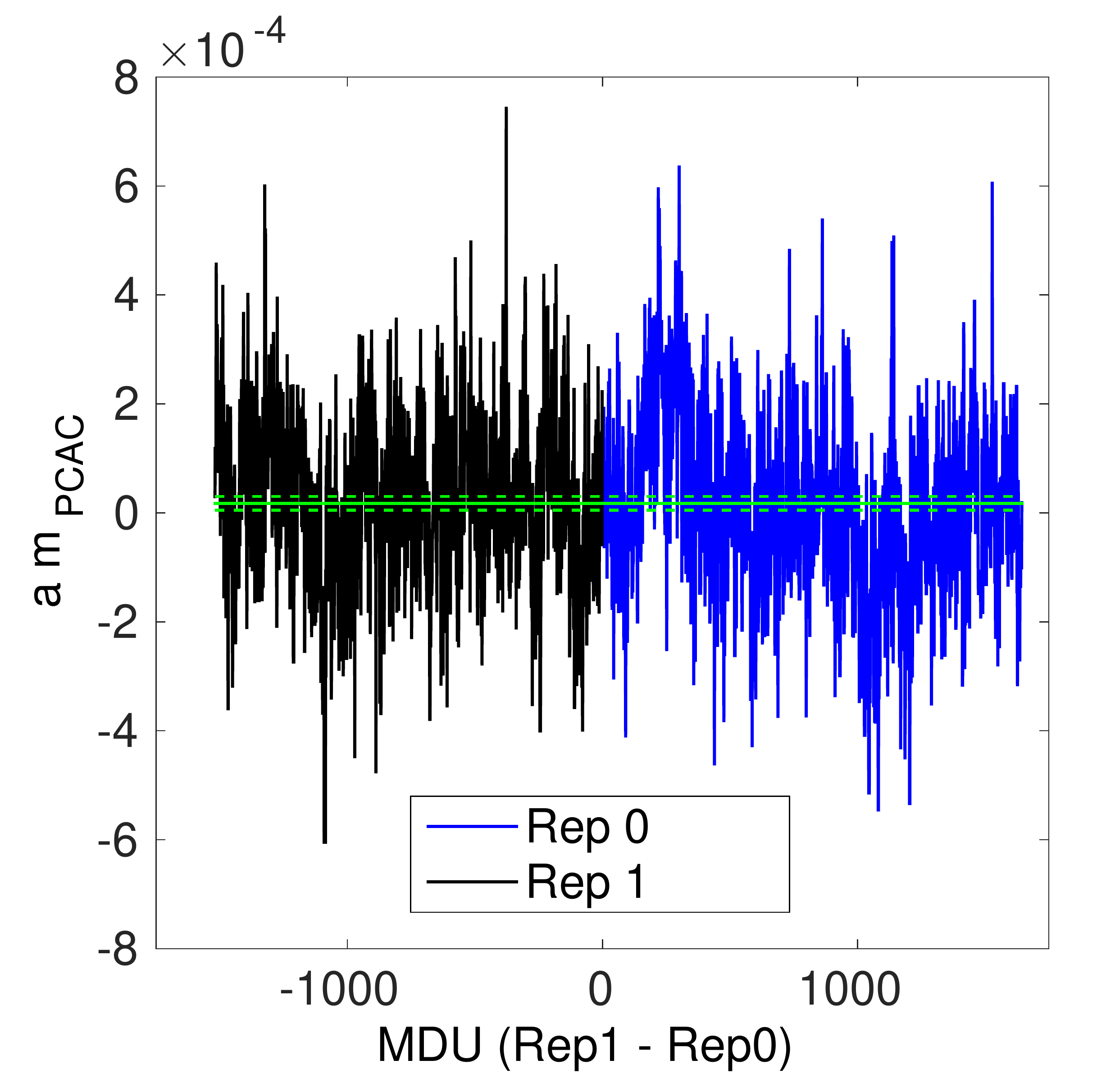}
\caption{\small Left: The PCAC mass versus the bare light quark mass $a m = 1/2\kappa - 4$
  for various values of the twisted mass parameter for the tuning of the critical mass. 
  The linear interpolations are done on the Th1.350.24 
ensembles (blue, triangle points) illustrated with the black solid line and 
on the Th2.200.32 ensembles (red square points) illustrated with the black dotted line. 
The value of the PCAC mass for the Th2.125.32.k1 ensemble is shown by the cyan star point
and of the cB211.072.64 by black right pointing triangle. 
Right: The MC History of the PCAC quark mass on the large volume physical 
point ensembles is shown at twisted mass value
$a \mu_\ell=0.00072$ and hopping parameter $\kappa=0.1394265$.}
\label{fig:tuning}
\end{figure}

For the tuning process of $\kappa$, which is a function of the 
light, strange and charm quark mass parameters, we use the 
$24^3\cdot 48$ and $32^3\cdot 64$ lattices, see section 3.2. for more 
details. The dependence  of the PCAC quark mass on $\kappa$ at fixed 
light twisted mass parameter is shown in Fig.~\ref{fig:tuning}. 
Note that it can be assumed that eq.~\eqref{eq:PCACslope} is valid here for the range $-0.4141 \lesssim a m_\ell \lesssim -0.4135$ 
i.e.~$|a(m_\ell - m_{crit})| < 0.0003$. Using simple linear fits for the $L=32$ ensembles, we determine a critical 
value of $\kappa$, $\kappa_{\rm crit} = 0.1394265$.
We then employ this $\kappa$-value for our large volume ensembles. 

For the simulations on the $64^3\cdot 128$ lattices we first thermalize 
one configuration using 500 trajectories. We then use this configuration 
as a starting point for two replicas, each having a final statistics 
of about 1500 MDUs. 
In Fig.~\ref{fig:tuning} we depict the Monte Carlo 
history of the PCAC quark 
mass for these two replicas, where we show, for better visibility, 
one history plotted by reversed history.
The PCAC quark mass fluctuates around zero and does not 
show particularly large autocorrelation times nor any indication 
of a first order Sharpe-Singleton transition. 
Performing the average over the two replica runs, 
we find $\mpcac/\mu= 0.03(2)$. 
Thus, the condition of eq.~(\ref{eq:condPCAC}) is nicely fulfilled. 
Note that here we do {\em not} include the 
renormalization factor $Z_A$. However, our first estimate is that $Z_A \approx 0.8$ and 
anyhow smaller than one, making the condition even better fulfilled.   
We therefore
conclude that the tuning to maximal 
twist is achieved for the
$N_f=2+1+1$ setup. And,  
as we will demonstrate below, the parameters of the cB211.072.64 runs 
are chosen such that we indeed simulate at, or very close to the physical 
values of the pion, the kaon and the D-meson masses.

\subsection{Tuning of the heavy quark sector}
\label{sse:thq}

\begin{figure}
\begin{center}
\includegraphics[width=0.49\textwidth]{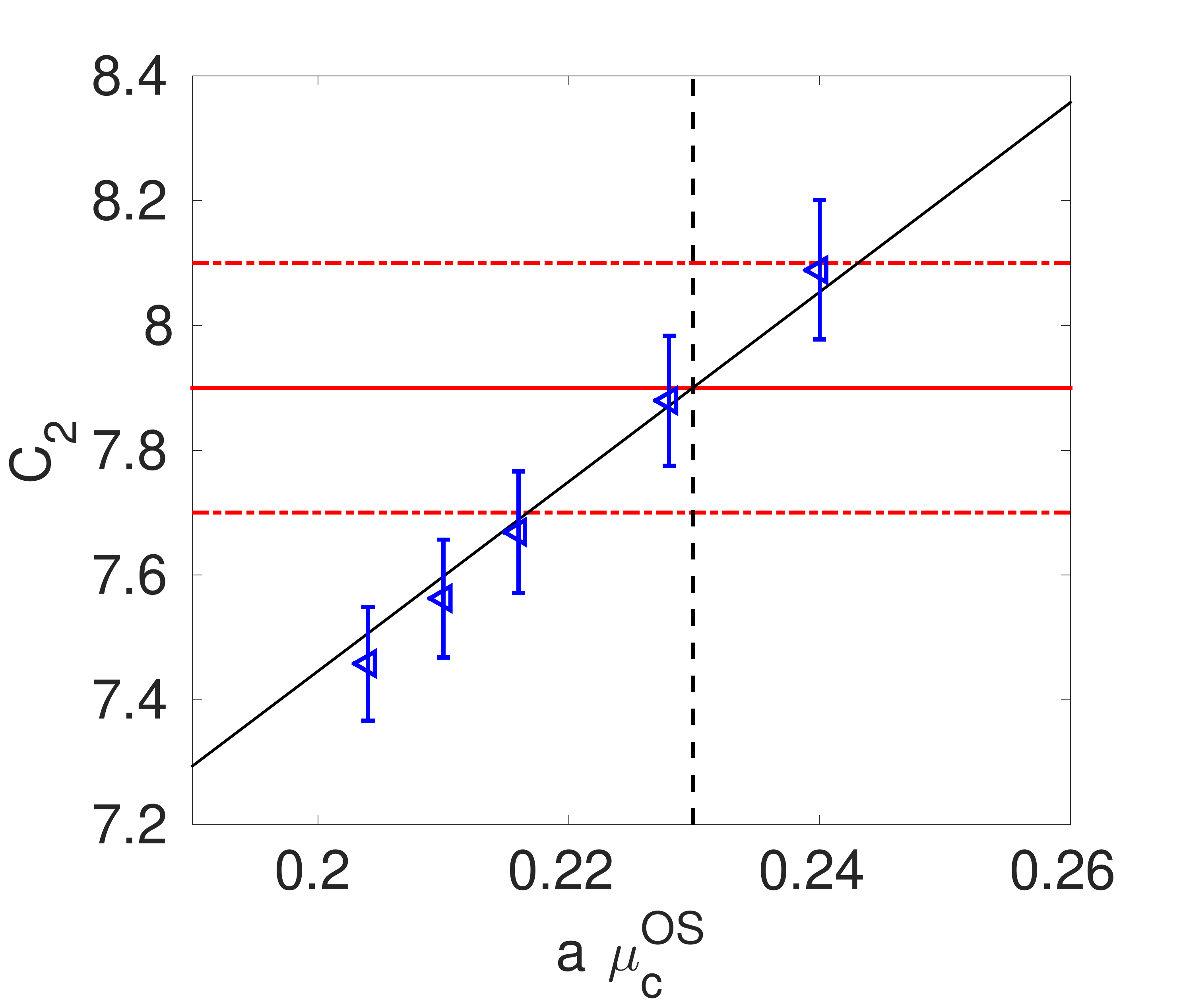}
\end{center}
\caption{\small Tuning of the charm quark twisted mass parameter $\mu_c$ using the $C_2$ condition on
the Th1.200.32.k2 ensemble.
The figure is showing a subset out of the 25 measured $m_{D_s}/f_{D_s}$ ratios, shown as the blue square points, using all combinations of 5 different
values of $\mu_c^{\rm OS}$ and $\mu_s^{\rm OS}$.
The horizontal line illustrate the physical value of $C_2$. The 25 points are interpolated and here showed
as the black straight line where $C_1$ is fixed.
}
\label{fig:Bandamu}
\end{figure}

In tuning the mass parameters of the heavy quark sector we exploit the fact  that 
the value of the critical hopping parameter, as determined
in the light quark sector, can be employed also for the 
heavy quark action while preserving automatic $\mathcal{O}(a)$-improvement
of all physical observables \cite{Frezzotti:2004wz,Chiarappa:2006ae}.
Nevertheless, tuning the heavy twisted mass parameters to reproduce the 
physical values of the strange and charm quark masses is a non-trivial task,
owing to the $\mathcal{O}(a^2)$ flavour violation~\cite{Frezzotti:2003xj} inherent to the 
heavy sector fermion action in eq.~(\ref{eq:sfheavy}). In order to tackle the problem, it
is convenient to employ in an intermediate step the so-called
Osterwalder Seiler (OS) fermions~\cite{Osterwalder:1977pc} in the valence which avoids these mixing effects. 
The OS-fermions can be 
used in a well defined mixed action setup as {\em valence} fermions at maximal 
twist with the same critical mass, $m_{crit}$, as determined in the unitary 
setup \cite{Frezzotti:2004wz}.
The flavour diagonal action, denoted as Osterwalder Seiler fermion action, is given by
\begin{equation}
S^{f}_{\rm OS} = \sum_{f=s,c} \left\{ \sum_x \bar\chi_f(x) \left[D_W[U] + \frac{i}{4} c_{SW} \sigma^{\mu\nu}
    \mathcal{F}^{\mu\nu}(U) + m_{crit} + i\mu_f^{\rm OS} \gamma_5 \right] \chi_f(x) \right\} 
\end{equation}
with $\chi_f$ a single-flavour fermion field. The renormalized
valence masses $\mu_{c,s}^{\rm OS,ren} = \mu_{c,s}^{\rm OS}/Z_P$ can be matched to the 
corresponding renormalized quark masses via
\begin{equation}
\label{eq:NDparcOS}
  \mu_{c,s}^{\rm OS,ren} = \frac{1}{Z_P} \left( \mu_\sigma \pm \frac{Z_P}{Z_S} \mu_\delta \right)  
\end{equation}
with $Z_P$ and $Z_S$ denoting the non-singlet pseudoscalar and scalar
Wilson fermion quark bilinear renormalization constants.
Then correlation functions using OS or unitary valence quarks are equivalent in the continuum.
Moreover they still yield $\mathcal{O}(a)$ improved physical observables.

The general idea to tune the heavy quark twisted mass 
parameters is to start with an educated guess in the 
unitary setup and to tune the OS charm and strange valence masses by imposing
two suitably chosen physical renormalization conditions.
The so determined parameters of the OS action, i.e.~$a \mu_s^{\rm OS}$ for the
strange quark and $a \mu_c^{\rm OS}$ for the charm quark, can then 
be translated to new heavy quark twisted mass parameters ($a \mu_\sigma$ 
and $a \mu_\delta$ of eq.~\eqref{eq:sfheavy}) via eq.~(\ref{eq:NDparcOS}), in 
the unitary setup and, together with a slight retuning of
$\kappa_{\rm crit}$, a new unitary simulation can be performed. 
With a convenient choice of the physical 
renormalization conditions, here $C_1$ and $C_2$ (see below), this parameter tuning procedure
can be carried out on a non-large lattice (in the present case, $32^3 \cdot 64$) 
and at larger than physical up/down quark mass.

In this work, we follow the above described strategy. 
As physical conditions we choose
\begin{equation}
C_1 \equiv  \frac{\mu_c^{\rm OS}}{\mu_s^{\rm OS}} = 11.8  
\qquad \textrm{and} \qquad  C_2 \equiv \frac{m_{D_s}}{f_{D_s}} = 7.9
\label{eq:tunc1}
\end{equation}
where $m_{D_s}$ is the $D_s$-meson mass and $f_{D_s}$ the $D_s$-meson decay constant.
The condition $C_2$ has a strong sensitivity 
to the charm quark mass while $C_1$ fixes the strange-to-charm mass ratio. 
They show only small residual light quark mass dependence arising from sea quark effects.
We expect these conditions to be essentially free from finite-size effects
due to the heavy $D_s$-meson mass.
This setup leads indeed to an only small error for the final parameter choices.
Details on our measurements of meson masses and decay constants 
for twisted mass fermions are given in the Appendix~\ref{ap:cor}. 

As a first step, we work on gauge ensembles produced with $\mu_\ell$ around three
times larger than the physical up-down average quark mass and with educated
guess values of $\mu_\sigma$, $\mu_\delta$ and $m_0$. We choose the OS 
quark masses $\mu_c^{\rm OS}$ and $\mu_s^{\rm OS}$ such that 
condition $C_1$ is fulfilled.
We then vary the OS quark masses, while maintaining condition $C_1$,
over a broad enough range such that also condition $C_2$ is satisfied 
within errors.

In a second step, we match the heavy charm and strange twisted mass of
the unitary action~(\ref{eq:sfheavy}) to the OS fermion quark mass parameters 
via eq.~\eqref{eq:NDparcOS}.
The value of $a \mu_\sigma$ is directly determined from
$a \mu_s^{\rm OS}$ and $a \mu_c^{\rm OS}$,
while $a \mu_\delta$ is fixed by the ratio $Z_P/Z_S$. The latter
can be estimated by adjusting $a \mu_\delta$  such that the kaon mass evaluated 
in the unitary formulation ($m_K^{tm}$) and its counterpart computed
with valence OS fermions ($m_K^{OS}$) are equal. 
Although the kaon mass value can be unphysical due to having a too large value of $\mu_\ell$ 
and possible finite size effects, the matching condition
actually relates only heavy quark action parameters. It fixes
the relation of $a \mu_\delta$ to $a \mu_s^{\rm OS}$ and $a \mu_c^{\rm OS}$,
or equivalently the ratio $Z_P/Z_S$.
In that way it is insensitive to both the finite lattice size and the actual value of
$\mu_\ell$ up to $\mathcal{O}(a^2)$ artifacts.
Since the matching steps described so far were implemented only on
the valence quark mass parameters of the unitary and OS actions using
gauge ensembles with so far different values of the sea quark mass parameters, 
one still needs to generate new gauge configurations at the so-determined 
values of $a \mu_\sigma$ and $a \mu_\delta$.
Now on these new ensembles it can be re-checked whether the
condition $C_2$ and the matching condition 
$m_K^{tm} = m_K^{OS}$, as well as the maximal twist condition eq.~\eqref{eq:condPCAC}
in the light quark sectors, are fulfilled with sufficient accuracy. 
If this happens not to be the case, the procedure has to be iterated.

More concretely, 
we start with an initial guess for the heavy quark mass parameters given by 
$a \mu_\delta = 0.1162$ and $a \mu_\sigma =0.1223$, which we deduce from a number of 
tuning runs on a lattice of size $24^3\times 48$ and 
$32^3\times 64$ along the lines of ref.~\cite{Abdel-Rehim:2014nka}. 
These parameters are realized for the ensemble 
Th1.200.32.k2, which is moreover very close to maximal twist.
We then employ OS fermions in the valence sector and 
vary the values of  $\mu_s^{\rm OS}$ 
and $\mu_c^{\rm OS}$ --while maintaining condition $C_1$--  
such that condition $C_2$ 
is fulfilled. This is illustrated in Fig.~\ref{fig:Bandamu} for the 
Th1.200.32.k2 ensemble. 
By requiring that condition $C_2$ is exactly fulfilled, we then fix       
the values of $a \mu_s^{\rm OS}$ and $a \mu_c^{\rm OS}$, finding 
\begin{equation}
 a \mu_s^{\rm OS} = 0.01948  \qquad \textrm{and}   \qquad a \mu_c^{\rm OS} =0.2299.  
 \label{eq:OSset1}
\end{equation}
 
As explained above, the values in eq.~(\ref{eq:OSset1}) already determine 
$a \mu_\sigma$. To determine $a \mu_\delta$, we first compute 
the kaon mass in the OS setup at $a \mu_\ell$ used in the unitary setup 
and $a \mu_s^{\rm OS}$ from eq.~(\ref{eq:OSset1}). 
Having found the OS kaon mass, we go back to the ensemble Th1.200.32.k2
and tune in the unitary heavy quark valence sector $\mu_\delta$ such that we match 
the OS kaon mass. We then take the so found value of 
$\mu_\sigma$ and $\mu_\delta$ for our simulations on the target 
large volume lattice.  
In this process a useful guidance is provided by assuming  
$Z_P/Z_S = 0.8$ known to be a typical value from our previous 
simulations. As we will discuss later, this assumption for 
$Z_P/Z_S$ turns out to be rather close to the values we 
determine on the cB211.072.64 ensembles. 
Our final result for the action parameters in the heavy sector of maximally 
twisted mass fermions then read 
\begin{equation}
a \mu_\sigma =0.12469 \qquad \textrm{and} \qquad a \mu_\delta =0.13151.
\label{eq:musigmadelta}
\end{equation}
Due to the retuning of the heavy quark masses $\kappa_{crit}$ has to be re-tuned
as well. To this end, several ensembles with volumes of $32^3\times 64$ at
light twisted mass values of $a \mu_\ell = 0.002$ and $a \mu_\ell = 0.00125$
were generated to determine the critical hopping parameter
for the simulation at $a \mu_\ell = 0.00072$ resulting in 
$\kappa_{crit} = 0.1394265$. 

In this work, it turned out that we only needed one iteration
of the above procedure using the Th1.200.32.k2 ensemble. 
After this first step, the tuning conditions for the heavy quark masses were checked again
on the Th2.200.32.k2 ensemble (see table \ref{tab:over}) and found to hold to a good accuracy
within statistical errors. A similar finding holds also on our target ensemble 
cB211.072.64 ensembles. If we impose again an exact matching between $m_K^{OS}$ and 
the unitary $m_K^{tm}$ on the two cB211.072.64 ensembles we find 
the ratio of the pseudoscalar to the scalar renormalization constants to be
\begin{equation}
 \frac{Z_P}{Z_S} = 0.813(1),
 \label{eq:ZPoZS}
\end{equation}
Using this value of $Z_P/Z_S$ the values of $\mu_{\sigma,\delta}$
of eq.~\eqref{eq:musigmadelta} are close to the corresponding parameters at the physical point 
(the cB211.072.64 ensembles) that match our tuning conditions.
Indeed the actually employed sea quark mass parameters correspond to a {\em sea} strange 
(charm) quark mass $6\%$ lighter ($4\%$ heavier) than those derived {\it a posteriori}
from imposing the same tuning and matching conditions on the physical point
ensembles. It is also very nice to observe that by enforcing these conditions with
very high precision one would obtain at the physical point with $a \sim 0.08$~fm
a kaon mass in isosymmetric QCD less than $1\%$ smaller than its experimental value.

\subsection{$\mathcal{O}(a^2)$ isospin-breaking lattice artifacts in the pion sector}
\label{sse:Isosp}

\begin{figure}
\begin{center}	
\includegraphics[width=\textwidth]{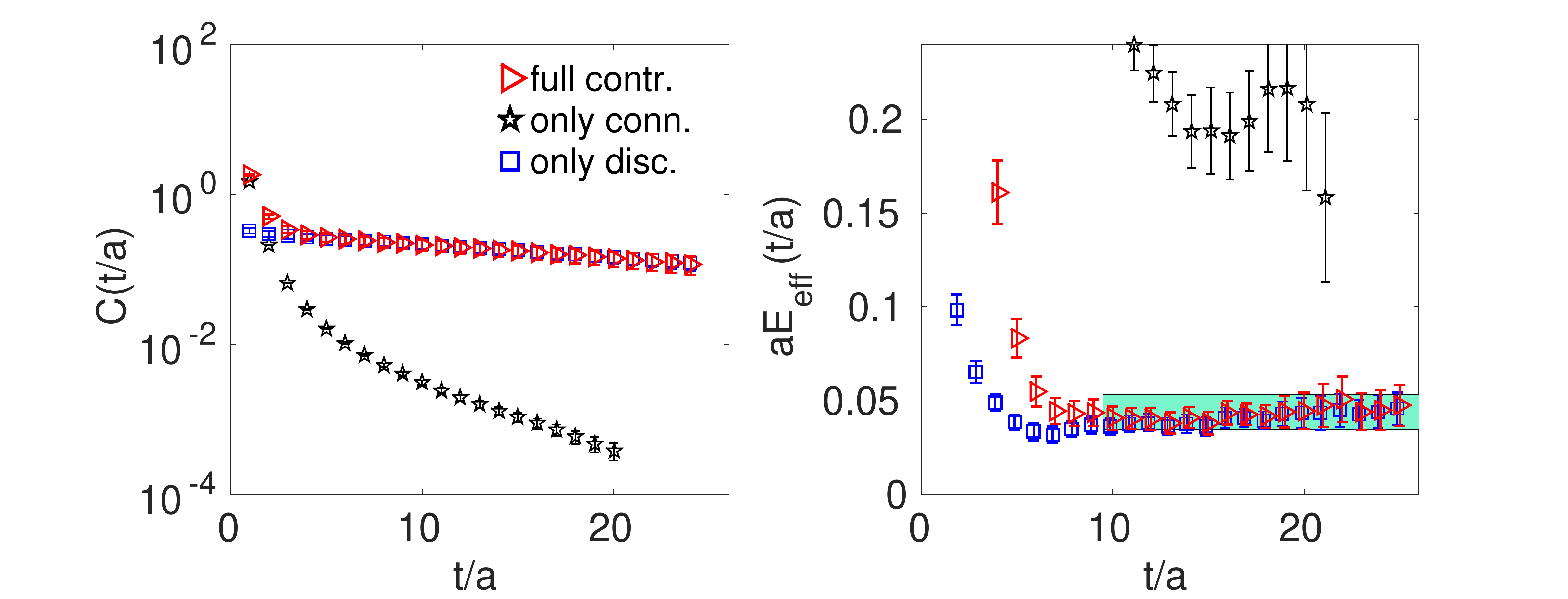}
\end{center}
  \caption{\small Left: The  correlator of the neutral pion versus $t/a$. 
    Right: The  effective mass of the neutral pion. The shaded band shows
    the constant fit in the plateau range. The red triangle shows the data of the full correlator,
    while the blue squares the disconnected and the the black stars the connected contribution.}
 \label{fig:pi0}
\end{figure}

An important aspect when working with twisted mass fermions 
at maximal twist is to keep the size of 
isospin violations small. 
This isospin breaking manifests itself by the fact that
the neutral pion mass becomes lighter than the one of the 
charged pion. In leading order (LO) of chiral perturbation theory this effect 
is described by 
\begin{equation}
 a^2 (m_{\pi}^2 - m_{\pi^0}^2) = - 4 c_2 a^2 \textrm{sin}^2 (\omega) 
 \label{eq:pionLOsplit}
\end{equation}
with the twisted mass angle given by $\omega = \textrm{atan}(\mu_\ell/Z_A \mpcac)$ and $c_2$ a low energy constant 
characterizing the strength of $\mathcal{O}(a^2)$-effects of twisted mass fermions. 
As shown in Ref.~\cite{Abdel-Rehim:2015pwa,Becirevic:2006ii}, 
using a clover term the value of the low energy constant $c_2$ decreases.
Indeed, employing a clover term, simulations at
physical quark masses become possible as demonstrated in Ref.~\cite{Abdel-Rehim:2015pwa}. 
It turns out that $c_2 < 0$ for twisted mass fermions~\cite{Herdoiza:2013sla} leading to the 
the Sharpe-Singleton scenario \cite{Sharpe:1998xm}. 

In order to calculate the neutral pion mass one needs to compute  disconnected
two-point functions that are 
notoriously noisy. To suppress the noise in the computation of the 
two-point functions we use
a combination of exact deflation, projecting out the 200 
lowest lying eigenvalues, and 6144 stochastic volume sources
corresponding to an eight-distance hierarchical probing~\cite{Stathopoulos:2013aci,Abdel-Rehim:2016pjw}.
The disconnected correlator needed is given by
\begin{equation}
 C_{disc}(t_0) = \langle \hat{O}(0) \hat{O}(t_0) \rangle \quad \textrm{with} \quad \hat{O}(t_0) = D^{-1}(t_0,t_0) - \langle D^{-1}(t,t) \rangle
\end{equation}
where the ensemble and time average of the vacuum contribution is subtracted from
the disconnected operator.
Note that we used global volume noise sources to extract the disconnected contribution, however
methods which do not subtract the vacuum expectation value explicitly could be more effective
as pointed out in \cite{Abdel-Rehim:2015pwa,Liu:2016cba,Ottnad:2017bjt}.
We have found that the disconnected contribution dominates the correlator for
time distances $t/a>10$, as can be seen in Fig.~\ref{fig:pi0}. However we include the 
connected contribution
in the plateau average,
leading to a neutral pion mass given by
\begin{equation}
    a m_{\pi^{(0)}} = 0.044(9)~.
\end{equation}
Note, that for the connected contribution small statistics of around 250 measurements is used,
which results in a relatively large statistical error.
The charged pion mass is straight forward to compute and we find for the charged pion mass $a m_\pi = 0.05658(6)$.
This gives an isospin splitting in the pion mass of $22(16)\%$ and the
low energy constant $c_2$ of eq.~\eqref{eq:pionLOsplit} reads
\begin{equation}
 4 c_2 a^2  = -0.0013(8)~
 \label{eq:c2}
\end{equation}
assuming $\omega = \pi/2$.
Thus, introducing a clover term for $N_f =2+1+1$ twisted mass fermions suppresses isospin breaking effects
effectively, i.e.~by a factor of $6$ compared to an $N_f=2+1+1$ ensembles with twisted mass fermions without a clover term and
a pion mass of 260 MeV at a similar lattice spacing of $a = 0.078(1)\;\textrm{fm}$~\cite{Herdoiza:2013sla,Baron:2010bv}, where
it was found that the mass splitting is given by $ (a m_{\pi^{(0)}})^2- (a m_\pi)^2 = -0.0077(4)$ .
The suppression of the pion isospin breaking effects, thanks to the use of the clover term,
is the underlying reason why we can perform our simulations at the 
physical point with $N_f=2+1+1$ flavours of quarks. 

\section{Pseudoscalar meson sector}
\label{sec:mesons}

In order to check, whether
we are indeed at (or close to)  the targeted physical situation, 
we studied the charged pion, the kaon and the D-meson masses and decay constants.
These observables are rather
straightforward to compute  with good
accuracy. A detailed description of the calculation 
of these quantities with twisted mass fermions can be found in appendix~\ref{ap:cor}~. 

\subsection{Light Meson sector}
\label{sec:lms}

\begin{figure}
\begin{center}	
  \includegraphics[width=0.49\textwidth]{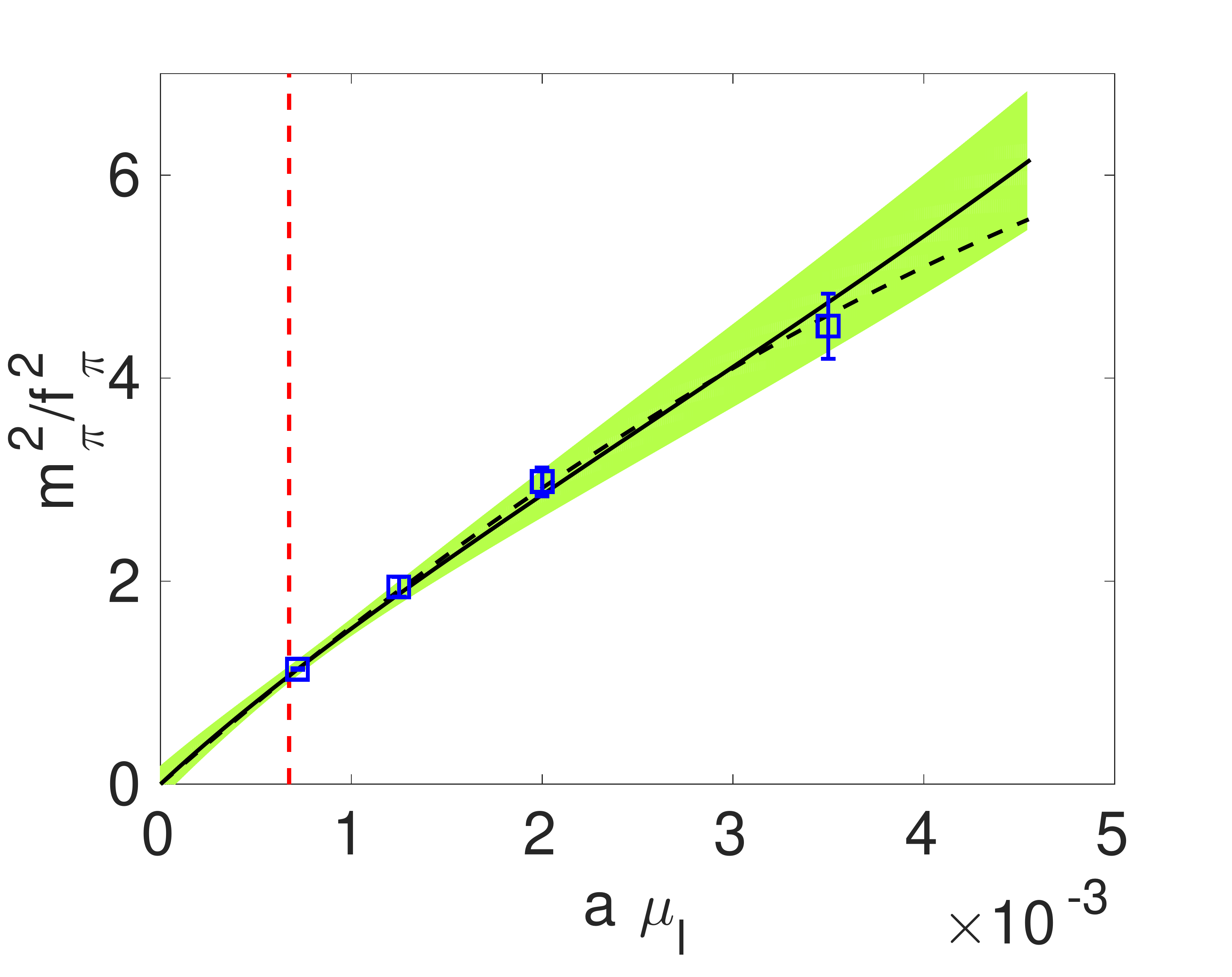}
\includegraphics[width=0.49\textwidth]{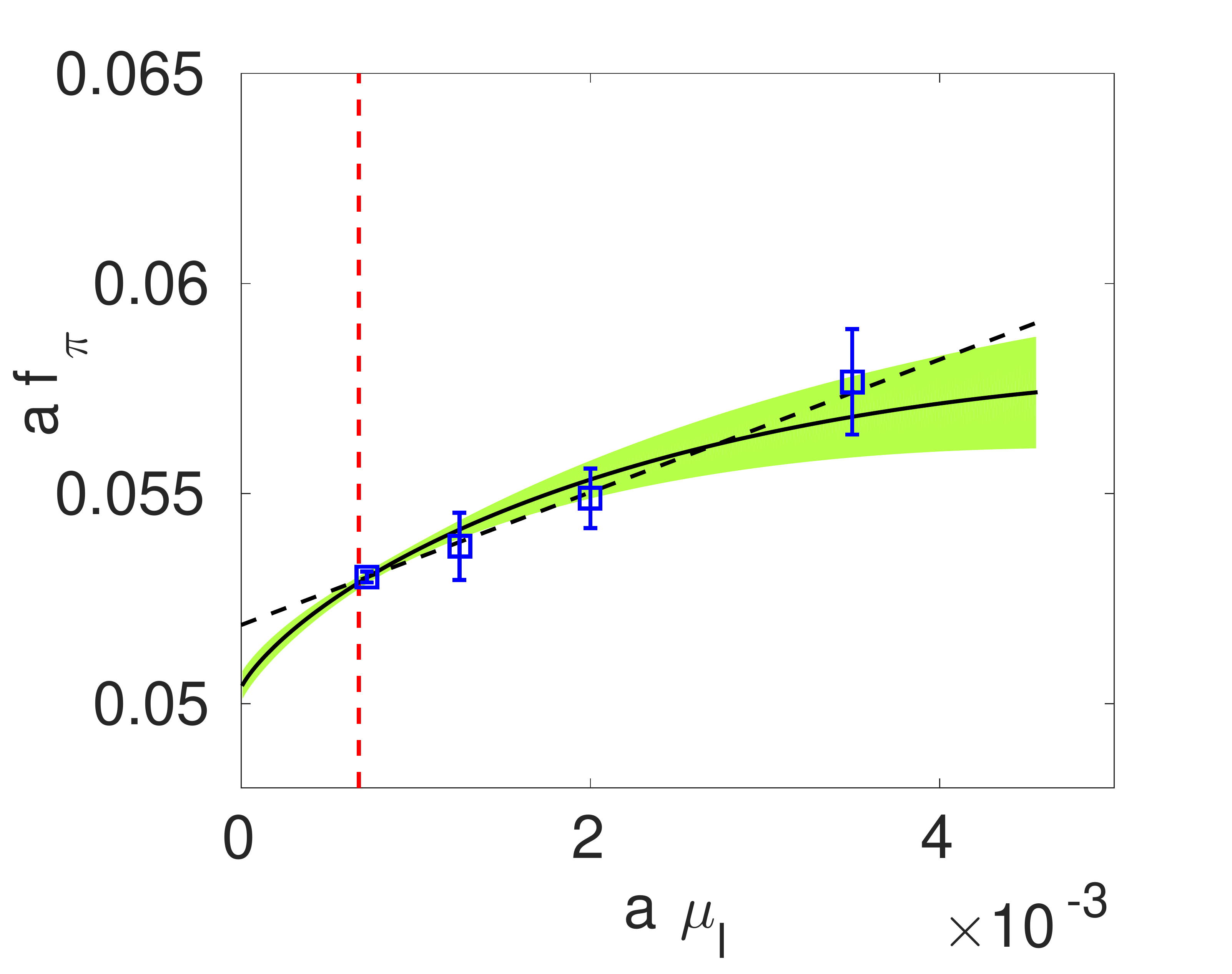}
\end{center}
  \caption{\small Left: The ratio $m_\pi^2 / f_{\pi}^2$ is plotted against $a \mu_\ell$.
  Right: The pion decay constant $a f_\pi$ is plotted against 
the light twisted mass math parameter $a \mu_\ell$. The solid lines are fits to 
NLO chiral perturbation theory with the error as shaded band, see eq.~(\ref{eq:pionmassoconst}) and 
eq.~(\ref{eq:pio}). The dotted lines are fits for which the chiral logs are neglected.
The pion mass and decay constant are corrected for by finite volume correction terms 
\cite{Colangelo:2005gd} and \cite{Colangelo:2010cu} respectively.
  }
 \label{fig:Mpi2ofpi}
\end{figure}

The first goal of this section is to determine the value of 
the lattice spacing within the pion sector. The extracted value
will then be compared to the one from a similar investigation in the 
nucleon sector in section~\ref{sec:bar}. 
In principle, the lattice spacing could be determined already 
from our cB211.072.64 target ensembles given in Table~\ref{tab:over}, having 
a twisted mass parameter
of $\mu_\ell = 0.00072$ and yielding a pion mass to decay constant ratio
of $m_\pi/f_\pi= 1.073(3)$, which is rather close to the 
physical one.
However, it is helpful to also use other ensembles,
listed in Table~\ref{tab:over}, which  
are all tuned to maximal twist, namely Th1.350.24.k2, Th2.200.32.k2, Th2.150.32.k2
in addition to the  cB211.072.64 ensembles. By employing chiral perturbation theory ($\chi$PT)  
to describe the quark mass 
dependence of the pion decay constant and pion mass, we obtain a
robust result for the value of the lattice spacing. 
Since the ensembles that are not at the physical point have 
partly only a small volume, we include finite volume corrections
from chiral perturbation theory to the $\chi$PT formulae used~\cite{Colangelo:2005gd}. 
We depict in Fig.~\ref{fig:Mpi2ofpi} 
the ratio $m_\pi^2/f_\pi^2$ and 
the pion decay constant itself as function of the light bare twisted 
quark mass.

In Fig.~\ref{fig:Mpi2ofpi} we also show the fits to 
NLO $\chi$PT~\cite{Weinberg:1978kz, Gasser:1983yg, Gasser:1984gg},  
which 
for the ratio $m_\pi^2/f_\pi^2$ read
\begin{equation}
 \frac{m_\pi^2}{f_\pi^2} = 16 \pi^2 \xi_\ell \left(1+ P \xi_\ell + 5 \xi_\ell \textrm{log}\left( \xi_\ell \right)\right) \frac{{F^{FVE}_{f_\pi}}^2}{{F^{FVE}_{m_\pi}}^2} 
\label{eq:pionmassoconst}
\end{equation}
and for the pion decay constant
\begin{equation}
 a f_\pi = a f_0 \left(1 + R \xi_\ell - 2 \xi_\ell \textrm{log}\left( \xi_\ell \right)\right) 1/F^{FVE}_{f_\pi} ,
\label{eq:pio}
\end{equation}
with the 
finite volume correction terms $F^{FVE}_{f_\pi}, F^{FVE}_{m_\pi}$~\cite{Colangelo:2005gd}.
Here  $\xi_\ell = 2 B_0 \mu_\ell / Z_P [(4 \pi f_0)^2]$ where $B_0$ and $f_0$ are low energy 
constants. From the fits, we determine the values of  
$2 B_0/Z_P = 4.52(6)$ and $a f_0 =  0.0502(3)$.  
The fitting constants $P,R$ are related to the NLO low energy constants by 
\begin{equation}
 P = -\overline{l}_3 - 4 \overline{l}_4 - 5 \textrm{log}\left(\frac{m_\pi^{phys}}{4\pi f_0}\right)^2 \quad \textrm{and} \quad R = 2 \overline{l}_4 + 2 \textrm{log}\left(\frac{m_\pi^{phys}}{4\pi f_0}\right)^2. 
\end{equation}
We determine the 
finite volume correction terms by fixing the low energy constants using the results of Ref.~\cite{Carrasco:2014cwa}.
For our target ensemble cB211.072.64 with $m_\pi L = 3.62$ we find that the finite volume effects 
yield corrections of less than $0.5\%$ for the pion mass
and less than $0.5\%$ for the pion decay constant.
By using the fit functions from $\chi$PT and fixing the  
ratio $m_{\pi,phys}^2/f_{\pi,phys}^2 \equiv 1.034$ we find for the light twisted mass parameter 
$a \mu_{\ell,phys} = 0.00067(1)$.
We then use this value in 
eq.~(\ref{eq:pio}) to determine
the lattice spacing. We get 
\begin{equation}
 a_{f_\pi} = 0.07986(15)(35) \; \textrm{fm}, 
\label{eq:lattmeson}
\end{equation}
with the first error the statistical and the second the systematic by using the physical value 
of the pion decay constant, $f_{\pi,phys} = 130.41(20)\; \textrm{MeV}$ \cite{Patrignani:2016xqp}.
We follow the procedure adopted in Ref.~\cite{Boucaud:2008xu} for determining
a systematic error by performing several different fits, adding or neglecting finite volume terms. Such fits employ 
e.g.~the finite volume corrections of 
Ref.~\cite{Colangelo:2010cu} using the calculated low energy constant $c_2$ of eq.~(\ref{eq:c2})
different orders in chiral perturbation theory and including or excluding the ensemble Th2.150.32.k2 due to
larger finite size effects. 
The systematic error is then given by the deviations of these different fits from the central value given in eq.~\eqref{eq:lattmeson}.
Although we include ensembles like Th2.150.32.k2 or Th1.350.24.k2 which have large finite size effect
of up to $8\%$ in the pion decay constant, the systematic uncertainties are suppressed due to the
fact that we are using ensembles close to physical quark masses which stabilize the fits.
Thus this demonstrates the importance of working at physical quark masses. 
Moreover this is confirmed by an estimation of the lattice spacing which takes
only the pion mass and decay constant from cB211.072.64 into account. 
Requiring a vanishing pion mass in the chiral limit,
the lattice spacing and the physical twisted mass value can be fixed by assuming a linear
dependence of $\mu$ on $a^2m_\pi^2$ and $m_\pi^2/f_\pi^2$. The so determined lattice spacing
agrees with eq.~(\ref{eq:lattmeson}) and reads $a=0.0801(2) \; \textrm{fm}$.
   
\subsection{Heavy meson sector}
\label{sec:twistedm}

As discussed in section~\ref{sse:thq}, the heavy sea quark
parameters used in the simulation are tuned by employing the ensemble Th1.200.32.k2.
With these parameters the kaon
mass on the cB211.072.64 ensembles is smaller as  compared to the OS kaon mass
using the parameters of eq.~(\ref{eq:OSset1}).
By employing the tuning condition of eq.~\eqref{eq:tunc1} 
we therefore re-adjust
the OS-parameters $a \mu^{\rm OS}_s$ and $a \mu^{\rm OS}_c$ following the tuning procedure of section \ref{sse:thq}, to take the values 
\begin{equation}
 a \mu_s^{\rm OS} =  0.01892(13) \quad \textrm{and} \quad a \mu_c^{\rm OS} = 0.2233(16)
\end{equation}
for the cB211.072.64 lattices. The OS valence quark parameters are lower by around 2.4\% compared to the
values determined using  the Th1.200.32.k2 ensemble (see eq.~\eqref{eq:OSset1}). By using $a \mu_{\ell,phys} = 0.000674$
the strange to light quark
mass ratio reads
\begin{equation}
 \frac{\mu_s^{\rm OS}}{\mu_l} =  \frac{0.01892(13)}{0.00067(1)} = 28.1(5)~.
\label{eq:heavyos}
\end{equation}

\begin{figure}
\begin{center}	
\includegraphics[width=0.45\textwidth]{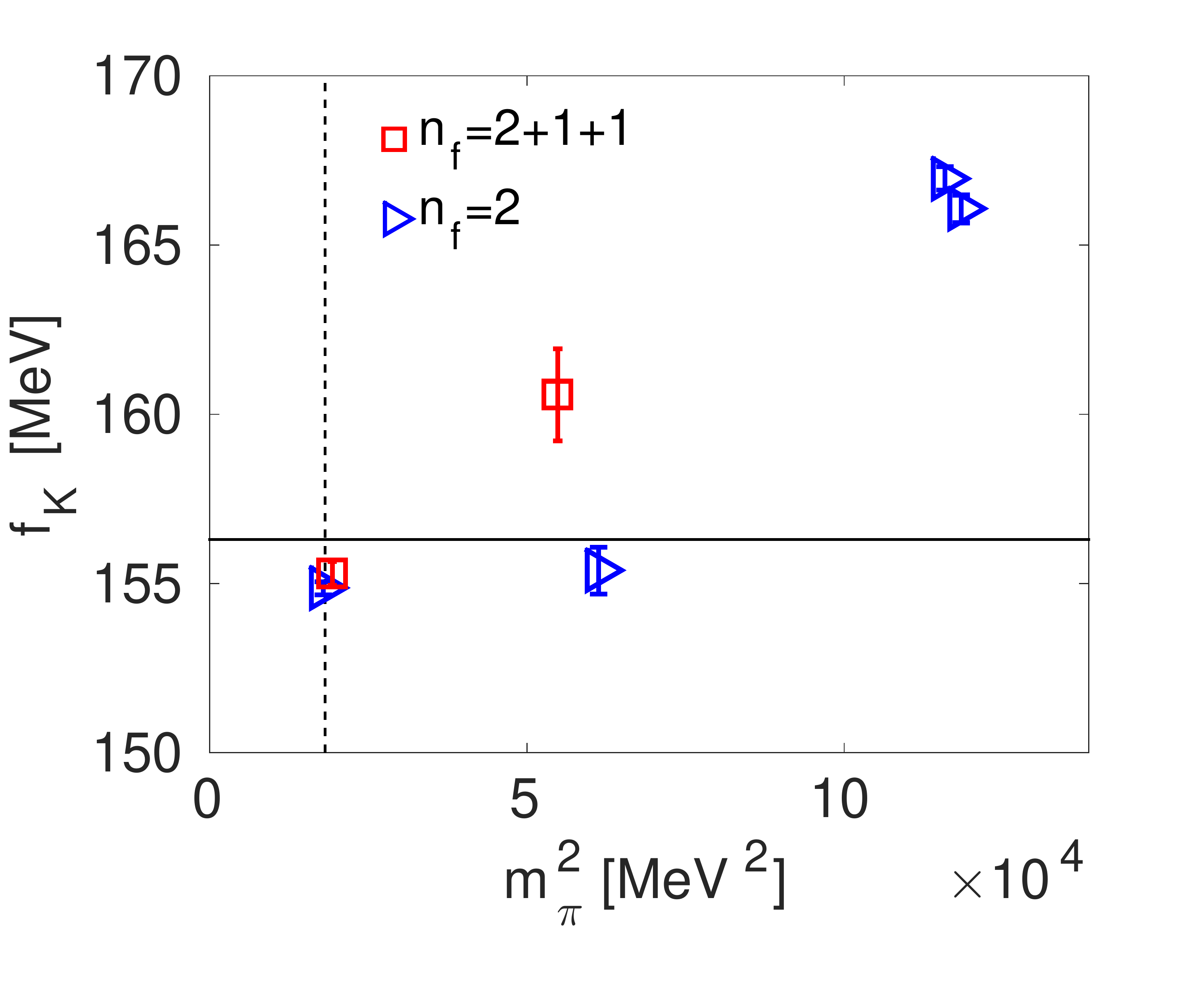}
\includegraphics[width=0.45\textwidth]{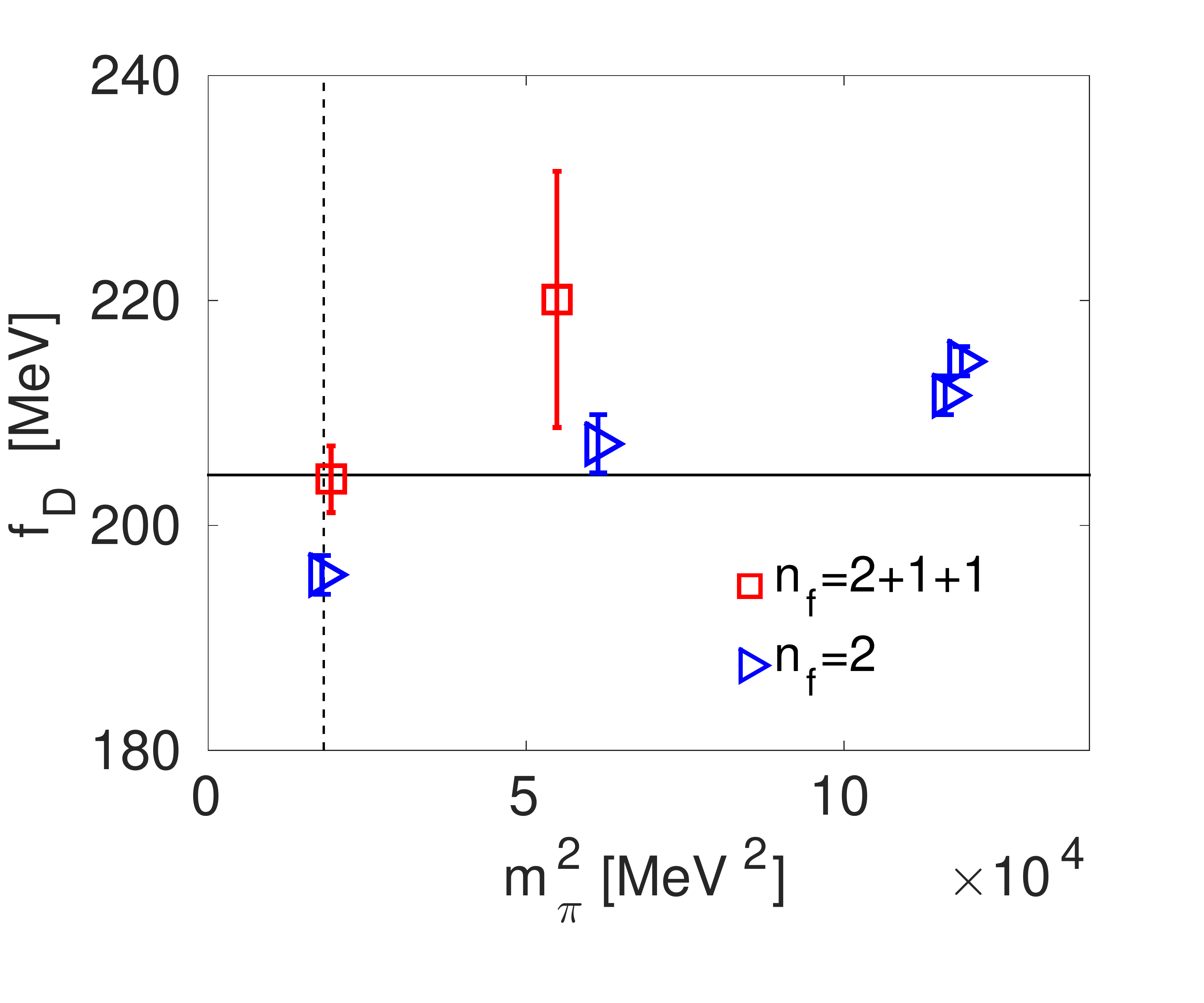}
\end{center}
  \caption{\small The pseudoscalar decay constants in the heavy quark sector. The left panel
  shows the kaon decay constant, while the right panel shows 
the D-meson decay constant both versus  the squared pion mass.
The dashed vertical line indicates the physical value of the pion mass. 
The red squares are  the measurements for the Th2.200.32.k1 and cB211.072.64 ensembles, while the blue
triangles are for the $N_f=2$ clover twisted mass ensembles~\cite{Abdel-Rehim:2015pwa}. The scale is set
via the pion decay constant.}
 \label{fig:fheavy}
\end{figure}

The kaon and D-meson masses and the respective decay constants as well as 
the corresponding quantities for the $D_s$-meson are all computed at three different values of $\mu_s^{\rm OS}$ and $\mu_c^{\rm OS}$.
We use a linear interpolation of $m_K^2$, $m_D$ and $m_{D_{s}}$ with respect to the heavy OS quark masses.
Using the values for $\mu_s^{\rm OS}$ and $\mu_c^{\rm OS}$ of eq.~\eqref{eq:heavyos} this allows us 
to determine  the masses and decay constants for these mesons. 
In  Fig.~\ref{fig:fheavy} we show the decay constants of the kaon and the D-meson 
and compare them with the results extracted from the $N_f=2$ clover ensembles~\cite{Abdel-Rehim:2015pwa}.
We employ 244 measurements for the cB211.072.64 and 100
for the  Th2.200.32.k1 ensemble.  The ratios of the kaon and D-meson masses to decay constants for the cB211.072.64 ensembles 
are found to be 
\begin{equation}
 \frac{m_K}{f_K} = 3.188(7) \qquad \textrm{and} \qquad \frac{m_D}{f_D} = 8.88(11),
\end{equation}
where the former ratio  has a central value slightly larger than the 
physical ratio 
$m_K^{\rm phys}/f_K^{\rm phys} = 3.162(18)$ \cite{Aoki:2016frl}, while the latter agrees well within errors
with the value $m_D^{\rm phys}/f_D^{\rm phys} = 9.11(22)$ \cite{Patrignani:2016xqp}.
These results indicate that discretization effects
for our setup 
are small in the heavy quark sector. For a more rigorous check,  a direct 
calculation at different values of the lattice spacing will be carried out. 

\begin{table}[thb]
  \centering
    \caption{The masses and the decay constants of the charged pseudoscalar mesons 
as well as the plaquette $P$ and $m_{\rm PCAC}$ are presented. }
  \begin{tabular}{|rl|rl|rl|}
     \hline 
     \vspace{-0.2cm} & & & & & \\
    $a m_\pi =$       &    0.05658(6)      &  $a m_K =$   &     0.2014(4) &  $a m_D = $     &  0.738(3)    \\[0.8ex]
    $m_\pi/f_\pi =$   &    1.0731(30)      &  $m_K/f_K =$   &   3.188(7) &  $m_D/f_D = $   &   8.88(11)  \\[0.8ex]
    \hline 
  \end{tabular}\label{tab:mesons}
  \begin{tabular}{|rl|}
     \hline 
     \vspace{-0.3cm}  &\\
   $a m_{\rm PCAC} =$       &    $0.189(114) 10^{-4}$  \\[0.8ex]
   $ P =$        &   $ 0.5543008(60)$  \\[0.8ex]
    \hline 
  \end{tabular}\label{tab:mesons}
\end{table}

\section{Baryon sector}
\label{sec:bar}

As another test, whether we are in the desired physical condition, we analyzed 
the nucleon mass which can also provide 
an independent determination of the lattice spacing, which can be 
compared to the one found in the meson sector. 
We measured the nucleon mass on the two cB211.072.64 ensembles
by using interpolating fields containing the operator
\begin{equation} 
  \label{eq:Proton}
  J_p = \epsilon_{abc}\bigl( u^T_a C\gamma_5 d_b\bigr)u_c\,,
\end{equation}
with $C=\gamma_4\gamma_2$ the charge conjugation matrix.
We then constructed the two point correlation function
\begin{equation}
  \label{eq:C_P}
  C_p(t) = \frac 1 2 {\rm Tr}(1 \pm \gamma_4) \sum_{\bf x}
  \langle J_p( {\bf x}, t) \bar J_p(0, 0)\rangle   
\end{equation}
which provides 
the nucleon mass in the large time limit. 
We used 50 APE smearing steps with $\alpha_{APE}=0.5$~\cite{Albanese:1987ds} in combination 
with 125 Gaussian smearing steps with $\alpha_{gauss} =0.2$~\cite{Gusken:1989qx, Alexandrou:1992ti} 
to enhance the overlap of the used point sources with the lowest state.

\begin{figure}
\begin{center}	
\includegraphics[width=0.49\textwidth]{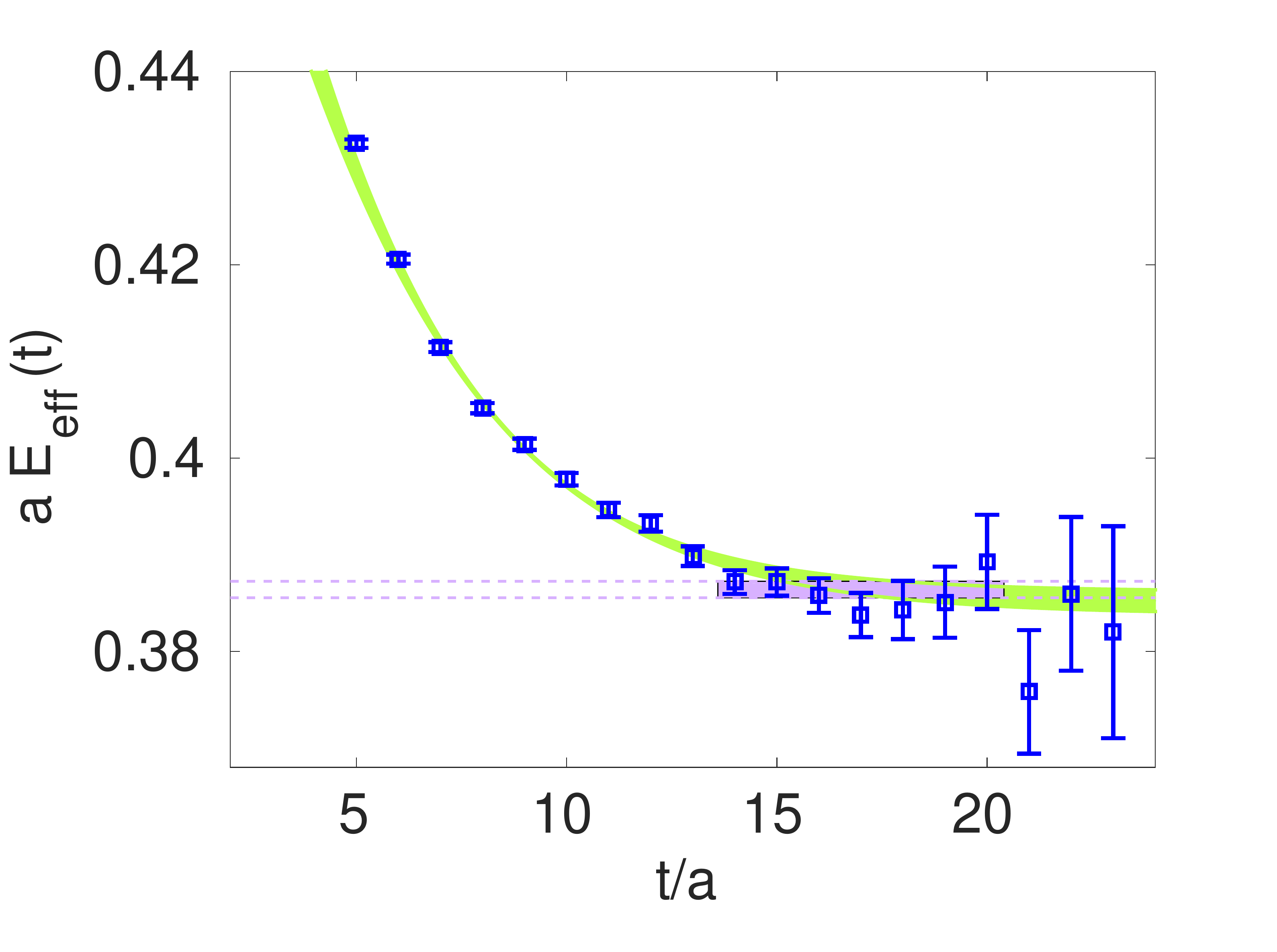}
\includegraphics[width=0.49\textwidth]{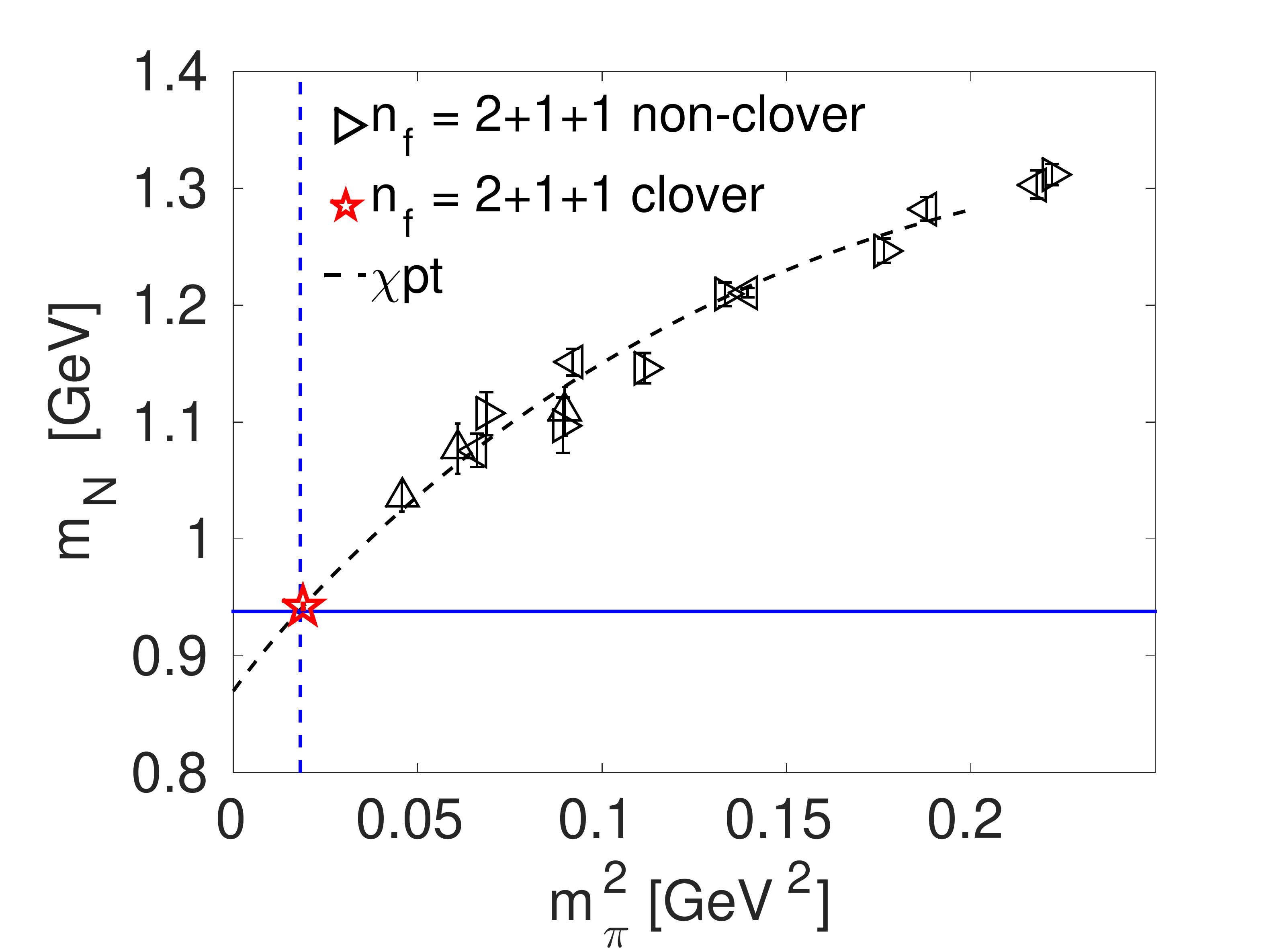}
\end{center}
  \caption{\small Left panel: The time dependence of the effective mass extracted from the nucleon correlator is 
  shown. The green shaded band corresponds to the two state fit while the violett band illustrate the used plateau average.
  Right panel: The squared pion mass dependence of the nucleon mass is shown by comparing
  the nucleon mass from our target lattice cB211.072.64 to the values determined on the $N_f = 2+1+1$ ETMC ensembles.
  The dotted line shows the fit by employing chiral perturbation theory at $\mathcal{O}(p^3)$.}
 \label{fig:nucleff}
\end{figure}

We extracted the nucleon mass for $t\gg 0$ by a plateau average over
the effective mass $a E_{eff} = \textrm{log}(C_p(t+a)/C_p(t))$ shown in Fig.~\ref{fig:nucleff}.
The plateau average of the nucleon mass, given by $am_N = 0.3864(9)$ 
on the cB211.072.64 ensemble, is in agreement
with a two-state fit with $am_{N,2st}= 0.3850(12)$ as shown in the left panel of Fig.~\ref{fig:nucleff}.

\subsection{Determination of the lattice spacing}
\label{sec:latsp}
As an alternative way to determine the lattice spacing, one can use the nucleon 
mass. 
A direct way would be to use the physical ratio
from which, by using the (lattice) pion mass
determined above, the lattice spacing can be estimated directly by the value
of the lattice nucleon mass.
Indeed, with the the pion mass $a m_{\pi} = 0.05658(6)$ the nucleon to pion mass ratio
$0.3864(9)/0.05658(6) = 6.83(2)$
is close to its physical value of $m_N^{\rm phys}/m_\pi^{\rm phys} = 0.9389/0.1348 = 6.965$
where we take the average of neutron and proton mass \cite{Patrignani:2016xqp}
and the pion mass in the isospin symmetric limit \cite{Aoki:2016frl}.
However, as in the case of the meson sector, using more data points
at heavier pion masses and $\chi$PT to describe their quark mass 
dependence, a more robust result can be obtained. 
More concretely, we have employed chiral perturbation theory 
at $\mathcal{O}(p^3)$ \cite{Gasser:1987rb,Tiburzi:2008bk}
for the nucleon mass dependence on the pion mass, i.e. 
\begin{equation}
 m_N = m_N^{phys} - 4 b_1 m^2_\pi - \frac{3 g^2_A}{16 \pi f^2_\pi} m^3_\pi~.
 \label{eq:mn}
\end{equation}
Similar to \cite{Alexandrou:2017xwd} we use the nucleon masses of the $N_f=2+1+1$ ETMC ensembles without
a clover term, determined in \cite{Alexandrou:2014sha} to perform 
the chiral fit of eq.~(\ref{eq:mn}). In our analysis we neglect cutoff effects,
which appear to be small and not visible within our statistical errors.
The same holds true for finite volume effects, see \cite{Alexandrou:2017xwd}. 
We fixed $f_\pi = 0.1304(2)\; \textrm{GeV}$  and $g_A = 1.2723(23)$ \cite{Patrignani:2016xqp} in 
eq.~(\ref{eq:mn}). 
The resulting fit to eq.~(\ref{eq:mn}) is shown in Fig.~\ref{fig:nucleff} (right) 
and allows to determine the lattice spacing as 
\begin{equation}
  a_{m_N}(\beta=1.778) = 0.08087(20)(37) \;\textrm{fm}~.
  \label{eq:latMN}
\end{equation}
The first error is statistical while the second error is the deviation
between the estimate obtained from eq.~(\ref{eq:mn}) and taken the mass from 
the two-state fit.
Note that the statistical error in the nucleon mass is comparable with pion sector due 
to three orders of magnitude larger number of inversions.

\subsubsection{$\mathcal{O}(a^2)$ isospin splitting in the baryon sector}

\begin{figure}
\begin{center}	
\includegraphics[width=0.65\textwidth]{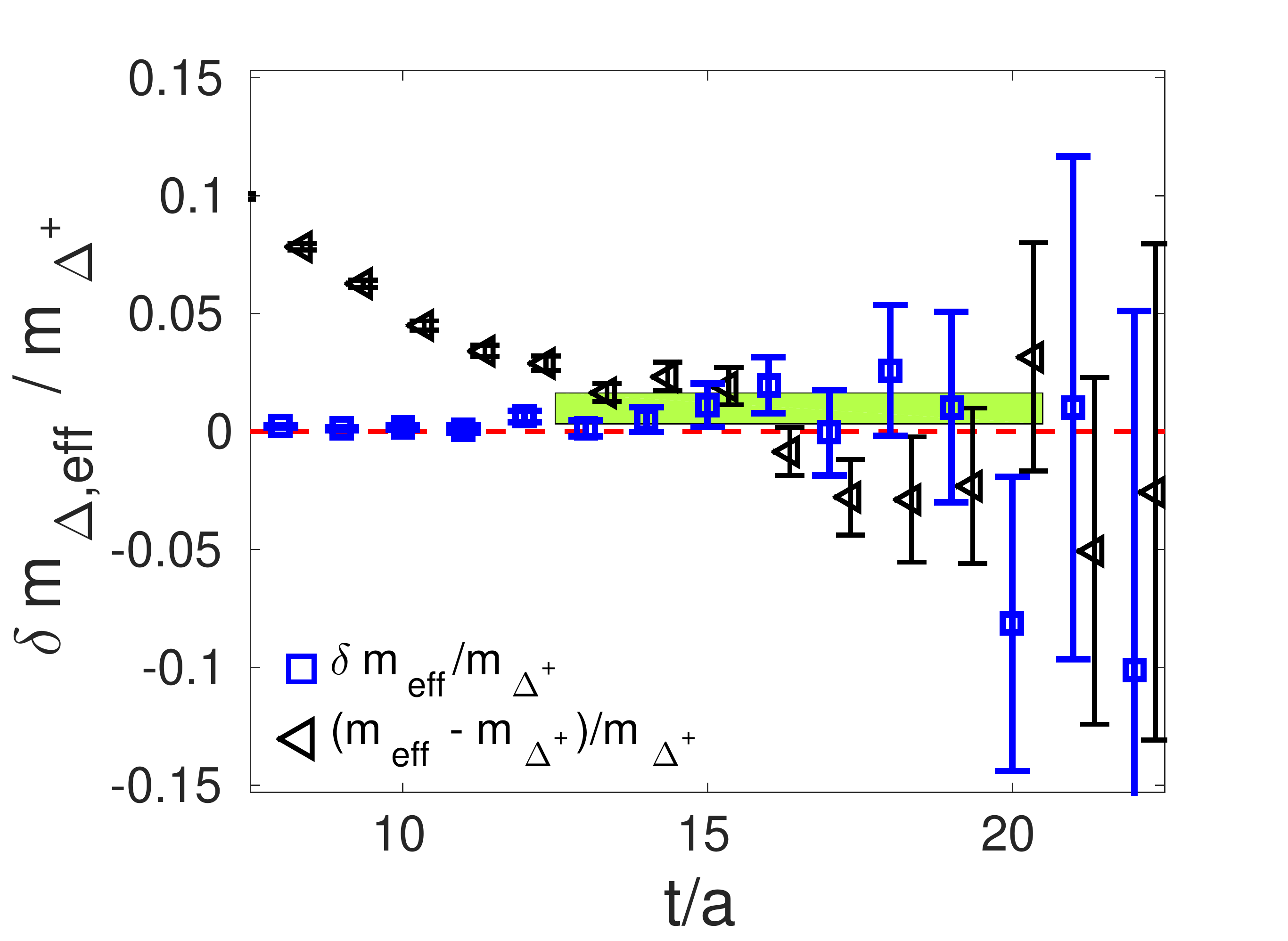}
\end{center}
  \caption{\small Relative differences of the effective Delta baryon masses: 
   The figure shows the relative difference given by $\delta m_{\Delta,eff}(t)/m_{\Delta^{+}}$ (blue squares). 
   To illustrate the beginning of the plateau we  added the relative effective mass $m_{eff}(t) - m_{eff}^{plateau}$ with $m_{eff}^{plateau}$
   is the plateau value $m_{\Delta^{+}}$ (black triangles).}
 \label{fig:deltaspli}
\end{figure}

The finite twisted mass value can result into a mass splitting of hadrons
which are symmetric under the isospin symmetry of the light flavor doublet.
As pointed out in sec.~\ref{sse:tlq} this indeed leads to a sizable effect
in the neutral-charged pion mass splitting. Here, we want to discuss the splitting
in the baryon sector in case of the $\Delta$-baryon employing the two cB211.072.64 ensembles.
Note that for the used lattice size the lowest decay channel of the Delta baryon,  
which is a nucleon+pion state with correct parity, is heavier than the Delta baryon itself. 
Thus, for the simulations performed here, the Delta can be treated as a stable state.

We measured the $\Delta$-baryon correlator by using the following interpolating fields
\begin{eqnarray}
  \label{eq:Delta}
  J^\mu_{\Delta^{+}} &=& \frac{1}{\sqrt{3}}\epsilon_{abc}\biggl[
  2\bigl( u^T_a C\gamma^\mu d_b\bigr)u_c +\bigl( u^T_a C\gamma^\mu u_b\bigr)d_c
  \biggl]~, \\
  J^\mu_{\Delta^{++}} &=& \epsilon_{abc}\bigl( u^T_a C\gamma^\mu u_b\bigr)u_c~.
\end{eqnarray}
Note that $J^\mu_{\Delta^{+}}$ and $J^\mu_{\Delta^{++}}$ is symmetric under $u\rightarrow d$ to $J^\mu_{\Delta^{0}}$
and $J^\mu_{\Delta^{-}}$ respectively.
We neglect the potential mixing of $\Delta$ with the spin-1/2 component
which is suppressed~\cite{Alexandrou:2008tn}. Thus the correlators
for the $\Delta^{++}$ is given by $C_{\Delta} = \textrm{Tr}[ C ]/3$
with $C_{ij} = \textrm{Tr}[(1+\gamma_4)/2 \langle J^i_{\Delta^{++}}(t) \bar{J}^j_{\Delta^{++}}(0)\rangle]$
and gives an average value of $a m_{\Delta} = 0.5251(72)$ by using a plateau average over the effective mass.
Now we define the splitting in the mass by
\begin{equation}
 \delta m_{\Delta,eff} = \textrm{log} \left\{ \frac{C_R(t)}{C_R(t+a)}\right\} \quad \textrm{with} \quad C_R = \frac{ C_{\Delta^{+}}(t) + C_{\Delta^{0}}(t) }{C_{\Delta^{++}}(t) + C_{\Delta^{-}}(t) }
\end{equation}
where we average over the symmetric parts.
In Fig.~\ref{fig:deltaspli} we show the effective relative mass splitting given by $\delta m_{\Delta,eff}/m_{\Delta^{+}}$.
In addition we plot the relative effective mass $m_{eff}(t)$ of the $\Delta^{+}$ particle subtracted from its plateau average
to illustrate where the plateau of the $\Delta$-baryon starts.
We find that the relative splitting in the $\Delta$ mass is  
$\delta m_{\Delta}/m_{\Delta^{+}} = 0.0098(65)$ and hence close to zero within errors.
This result is in agreement with \cite{Abdel-Rehim:2015pwa}
where it was found that the isospin splitting of the twisted mass action 
in the baryon section is suppressed.

\section{Lattice spacing}
\label{sec:olatsp}

\begin{figure}
\begin{center}	
  \includegraphics[width=0.49\textwidth]{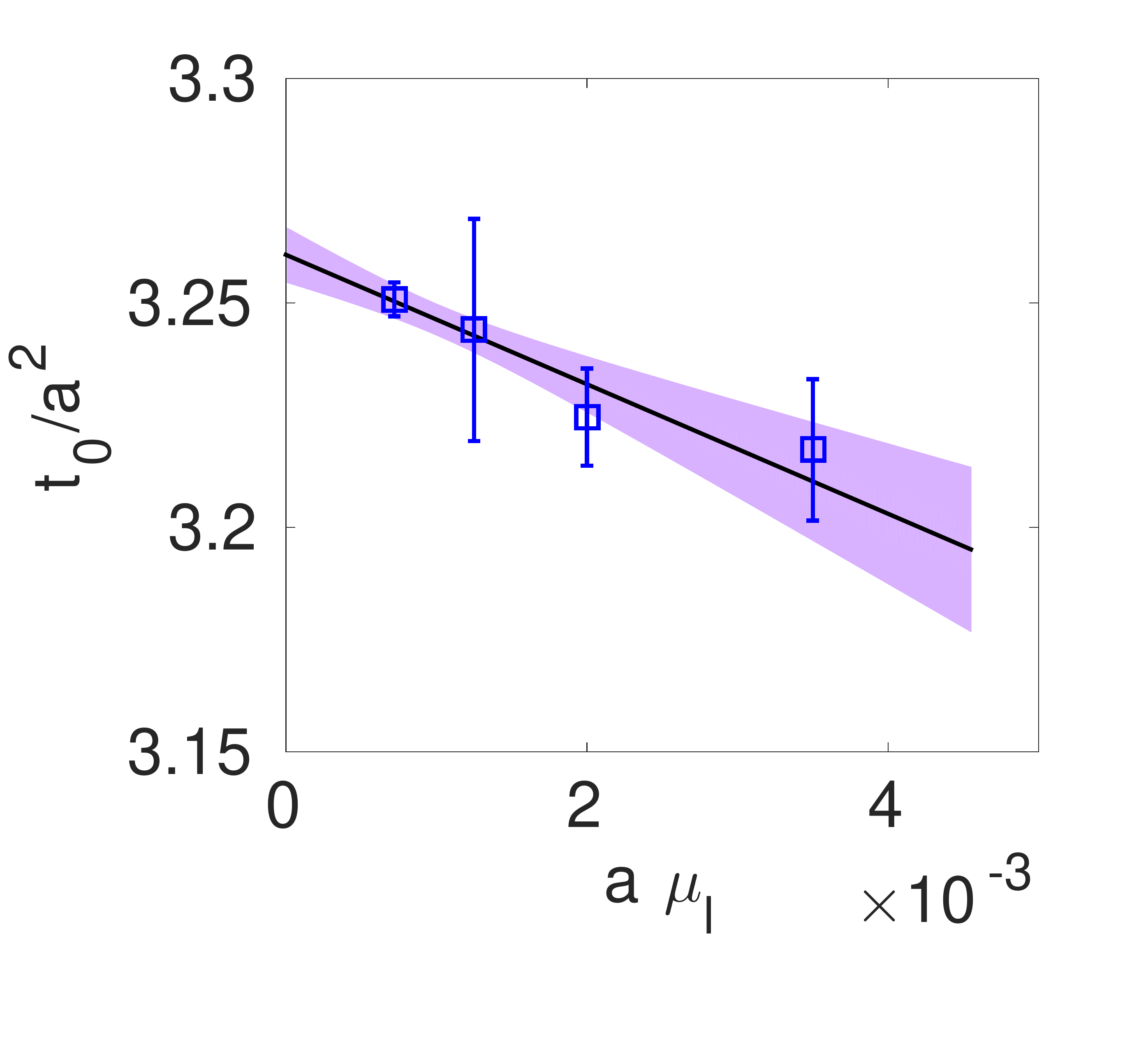}
\includegraphics[width=0.49\textwidth]{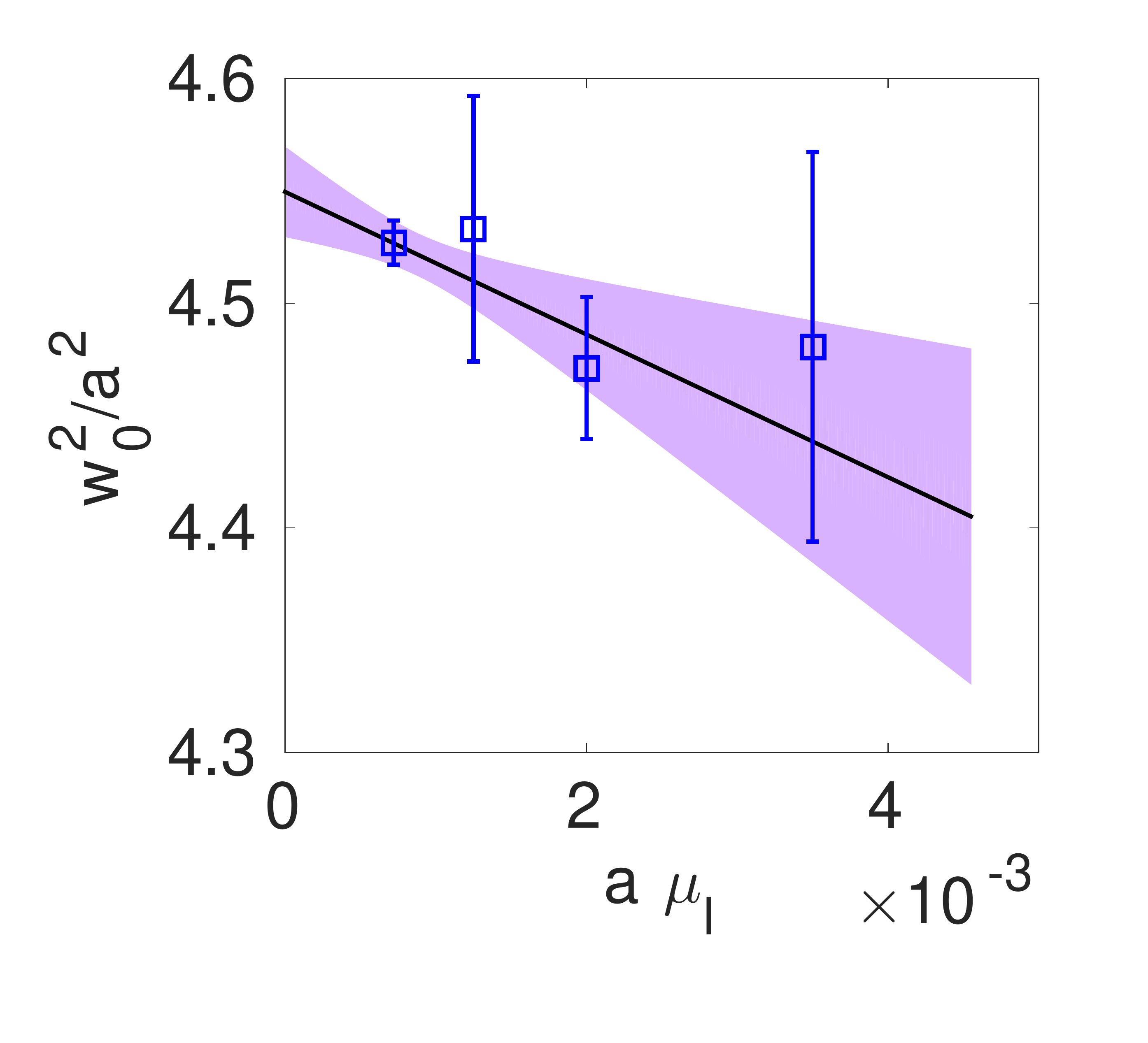}
\end{center}
  \caption{\small Left: Linear extrapolation of the gradient flow observable
  $t_0/a^2$. Right: Linear extrapolation of the gradient flow observable
  $w_0^2/a^2$. The solid line with the shaded violett band shows the linear extrapolation. 
  }
 \label{fig:gradflow}
\end{figure}

The lattice spacing can be evaluated  by matching lattice observables
to their physical counterparts. This has been 
done, as described in sections~\ref{sec:lms} and \ref{sec:latsp}, in the meson sector by employing the pion decay
constant  and in the baryonic sector using the nucleon 
mass, respectively.
Differences in the values obtained for the lattice spacing as determined using different physical observables 
can shed light on cut-off effects.
We  discuss  in this section
an additional method to determine the lattice spacing,
which is provided by the gradient flow scale setting parameters 
$t_0$~\cite{Luscher:2010iy} and $w_0$ \cite{Borsanyi:2012zs}. 
Following the procedure described in these articles and 
in particular as applied to the twisted mass setup~\cite{Abdel-Rehim:2015pwa}, 
we extrapolate the gradient flow observables to the chiral limit 
using a fit ansatz linear in $a\mu_\ell$, which corresponds to LO $\chi$pt \cite{Bar:2013ora}.
The resulting curve is shown as Fig.~\ref{fig:gradflow}. We follow a similar procedure for the extrapolation of $w_0^2/a^2$.
We employ the values computed for the ensembles Th1.350.24.k2, Th2.200.32.k2, Th2.125.32.k1,
cB211.072.64 and find $t^{\textrm{ch}}_0/a^2= 3.261(6)$ and ${w^{\textrm{ch}}_0}^2/a^2= 4.550(20)$.
Using the phenomenological values of 
$\sqrt{t_0}=0.1465(25)$ and $w_0=0.1755(18)$~\cite{Borsanyi:2012zs}
we deduce the following values for the lattice spacing
\begin{equation}
 a_{t_0} = 0.0811(14)\;\textrm{fm} \qquad \textrm{and} \qquad a_{w_0} = 0.0823(8) \; \textrm{fm}~.
\end{equation}

In Table~\ref{tab:latsp} we summarize the values of the lattice spacing
as determined from the
pion mass and decay constant, the nucleon mass and the gradient flow 
parameters $t_0$ and $w_0$. As it can be noticed,  
there are small deviations of the lattice spacing 
between the meson and the baryons sector and in any case they are
comparable to the one we have observed in the simulations 
with $N_f=2$ flavours of quarks. That indicates that cutoff effects
do not increase for our $N_f=2+1+1$ flavour setup used here. 
We would like to stress, that we plan to
carry out further simulations at different and, in particular, smaller values of the 
lattice spacing in future works.

\begin{table}[thb]
  \centering
    \caption{
    We give the values of the lattice spacing determined by using different physical
quantities as inputs, including in the errors the input systematic uncertainties. The final value of the lattice spacing is derived via a 
weighted average of $a_{f_{\pi}}$  and $a_{m_N} $  where for the final error a $100\%$ correlated data is assumed \cite{1402-4896-51-6-002}. The residual 
systematic uncertainty on the lattice spacing, which stems from higher order cutoff effects,
should be of relative size $\mathcal{O}(a^2)$ and looks numerically smaller than $2\%$.}
\begin{tabular}{||c|l||rl||}
     \hline 
          \vspace{-0.3cm} & & & \\
     phys.~quant. & lat.~spac. [fm] &  quantities in lat.~units &  \\
     \hline 
     \vspace{-0.2cm} & & & \\
   $  a_{t_0}     $      &    0.0811(14)      & $\left.t_0/a^2\right|_{\mu_\ell=0.00072}=  $   &   $3.246(\phantom{0}7)$   \\[0.8ex]
   $  a_{w_0}     $      &    $0.0823(\phantom{0}8)$      & $\left.w_0^2/a^2\right|_{\mu_\ell=0.00072}=$   &   4.512(16)  \\[0.8ex]
   \hline
   \hline
   \vspace{-0.2cm} & & & \\
   $a_{f_{\pi}}    $       &    0.07986(38)       & $\left.af_\pi\right|_{\mu_\ell=0.00072}=$    &   0.05272(10)  \\[0.8ex]  
   $a_{m_N} $            &    0.08087(44)     & $\left.m_N/m_\pi\right|_{\mu_\ell=0.00072}=$   &   6.829(19)  \\[0.8ex]
 
\hline 
\vspace{-0.2cm} & & & \\
    average       &   0.08029(41)     &   &   \\[0.8ex]
   \hline 
  \end{tabular}\label{tab:latsp}
\end{table}

\section{Conclusions\label{sec:conclusions}}

The first successful simulation 
of maximally twisted mass fermions with $N_f=2+1+1$ quark flavours at 
the physical values of the pion, the kaon and the D-meson masses has been presented. 
By having a lattice spacing of $a=0.08029(41) \;\textrm{fm}$, we find that the simulations 
are stable when performed with physical values of the quark mass parameters. In particular, we are able
to carry out a demanding but smooth tuning procedure to maximal 
twist and to find the values
of the light, strange and charm bare quark masses, which correspond to 
the physical ones for the first two quark generations. 

In our setup, which employs a clover term, the cutoff effects appear 
to be small. Several observations corroborate  this conclusion: as already 
mentioned above, the simulations themselves are very stable; when fixing 
the quark mass parameters through the selected physical observables, other
physical quantities, as collected in Table~\ref{tab:mesons} come out to be 
consistent with their physical counterparts; the $\mathcal{O}(a^2)$ effects originating 
from the isospin breaking of twisted mass fermions are small and significantly 
reduced compared to our earlier simulations with $N_f=2+1+1$ flavours at 
non-physical pion masses; deviations of the lattice spacing from the 
meson sector, the baryon sector and gradient flow observables, as listed 
in Table~\ref{tab:latsp}, are small and of the same size as in our former 
$N_f=2$ flavour simulations. 

This work focuses on the tuning procedure both to maximal twist and 
to the physical values of the quark masses. We include the  pseudoscalar meson
masses and decay constants as well as the nucleon and $\Delta$ masses, in order
to demonstrate that we indeed reach the targeted physical setup. We are planning
to compute many more quantities in the future connected to hadron structure, scattering phenomena, 
electroweak observables and heavy quark decay amplitudes. In addition, we 
have already performed the tuning for a second, finer lattice spacing and 
we are in the process of generating configurations. 
The combination of results for various physical quantities 
from the present lattice spacing of $a\approx 0.08\;{\rm fm}$,   
from the ongoing finer lattice spacing and from an already existing 
lattice spacing of $a\approx 0.1\;{\rm fm}$, which is however not exactly at the physical point, will 
allow us 
to explicitly check the size of cut-off effects and eventually take the 
continuum limit. 

We thus conclude that we have given a successful demonstration that simulations 
of maximally twisted mass fermions with $N_f=2+1+1$ quark flavours 
can be carried out with all quarks of the first 
two generations tuned to their physical values.  
This clearly opens the path for the ETM collaboration to perform simulations 
towards the continuum limit with a
rich research programme being relevant for phenomenology and ongoing and 
planned experiments.

\subsection*{Acknowledgments}
We would like to thank all members of the ETM Collaboration for a productive collaboration. 
This project has received funding from the  Horizon 2020 research and innovation programme of the European Commission
under the Marie Sklodowska-Curie grant agreement \emph{No 642069}. S.B.~is supported by this programme.
The authors gratefully acknowledge the Gauss Centre for Supercomputing e.V.~(www.gauss-centre.eu)  
for  funding  the project \emph{pr74yo} by  providing  computing  time  on  the  GCS  Supercomputer 
SuperMUC at Leibniz Supercomputing Centre (www.lrz.de), where the main simulations were performed.
Part of the results were obtained using Piz Daint at Centro Svizzero di Calcolo Scientifico (CSCS),
via project with id \emph{s702}.  We thank the staff of LRZ and CSCS for access to the computational resources
and for their constant support as well as the Julich Supercomputing  Centre  (JSC)  for  the  tape  storage.
Part of this work was supported by the DFG Sino-German \emph{CRC110}.
\appendix
\section{Mesonic Correlators}
\label{ap:cor}

In general, the charged 2-point pseudoscalar correlators can be defined by
\begin{equation}
  \label{eq:Cps}
  C_\mathrm{PS}^{q,q'}(t)\ =\ \langle\, P^\pm_{q,q'}(t)\ P^\pm_{q,q'}(0)^\dagger\, \rangle
\end{equation}
using the interpolating field
\begin{equation}
  P^\pm_{q,q'}(t)\ =\ \sum_\mathbf{x}\bar{\chi}_q(\mathbf{x},t)\, i\gamma_5\,
  \tau^\pm\, \chi_{q'}(\mathbf{x},t)\,,\qquad \tau^\pm =
  \frac{\tau^1\pm i\tau^2}{2}
  \label{eq:PSinter}
\end{equation}
with the quark flavors $q,q'\in \{\ell,s,c\}$. 
For sufficiently large times the charged pseudoscalar correlator is dominated by the lowest energy, such that 
\begin{equation}
    C(t)_{ ~ \overrightarrow{t  \gg a, ~ (T - t) \gg a} ~ } \frac{G_{PS}^2}{2m_{PS}} \left( e^{ - m_{PS}  t}  + e^{ - m_{PS}  (T - t)} \right) 
    \label{eq:larget}
\end{equation}
and the mass $m_{PS}$ and matrix element $G_{PS} = | \langle 1_{PS} | P^\pm_{q,q'} | 0 \rangle|$ can be extracted 
in a standard way 
via plateau averages for large time extent.
In case of maximal twist the matrix element $G_{PS}$ is directly 
connected to the pseudoscalar decay constant by \cite{Frezzotti:2000nk,Frezzotti:2003ni}
\begin{equation}
     f_{\pi} =   \frac{(\mu_q + \mu_{q'}) G_{PS}}{\textrm{sinh}(m_{PS}) m_{PS} } ~ .
    \label{eq:decaypi}
\end{equation}
Due to the flavor mixing in case of the mass non-degenerate 
twisted mass operator we adopt a non-unitary setup \cite{Frezzotti:2004wz} for
the heavy quark doublet, namely the Osterwalder-Seiler fermion regularization \cite{Osterwalder:1977pc}.
As shown in \cite{Frezzotti:2004wz}, 
this mixed action introduces effects which are only of order ${\cal{O}}(a^2)$
and are hence suppressed for small $\mu$ and fine lattice spacings.
The OS fermions correspond to the twisted mass discretization in 
single flavor space, where $\mu = \pm \mu_q$.
The sign of $\mu$ is always chosen such that the two valence quarks 
in the interpolating fields eq.~\eqref{eq:PSinter} have opposite signs.

\section{Autocorrelation}
\label{ap:au}

\begin{figure}
\begin{center}	
\includegraphics[width=\textwidth]{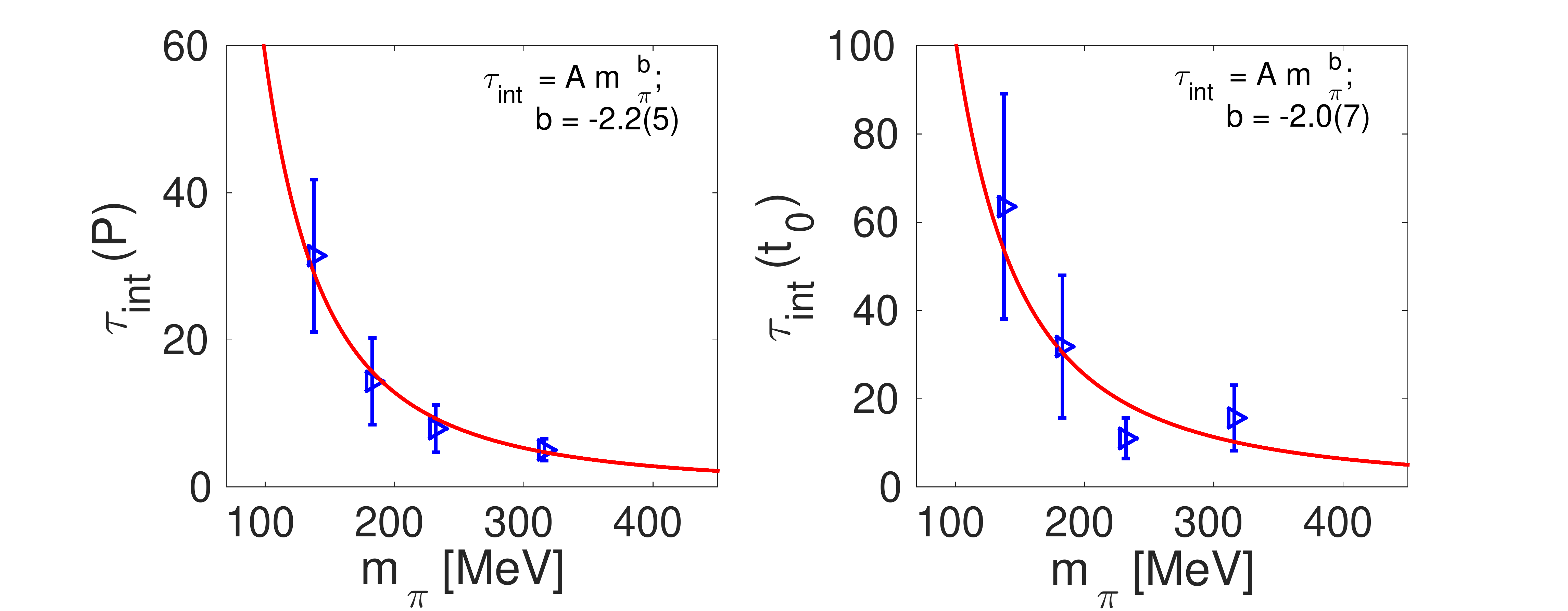}
\end{center}
  \caption{\small The figure shows the integrated autocorrelation $\tau_{int}$ of the ensembles at maximal twist (from right to 
  left Th1.350.24.k1, Th2.200.32.k2, Th2.125.32.k1, cB211.072.64 for both cases). Note that we used here the ensemble Th1.350.24.k1
  instead of Th1.350.24.k2 which not at maximal twist but has a larger statistics.}
 \label{fig:autoc}
\end{figure}

The autocorrelation of the Hybrid Monte Carlo algorithm slows down critically 
for very fine lattice spacings with $a<0.05 \; \textrm{fm}$. 
This can be seen in the freezing of the topological charge \cite{Schaefer:2010hu}.
For our lattice with $a\sim 0.08\; \textrm{fm}$ we found that the 
topological charge can fluctuate between the different sectors leading to 
small autocorrelation times of $\tau_{int}(Q)=13(5)\; \textrm{[MDU]}$. 
As pointed out in \cite{Bruno:2014ova} the energy density at finite flow times develops larger autocorrelation times
in the regime with $a \gtrsim 0.05 \; \textrm{fm}$.
Although we have relative small statistics we calculated integrated 
autocorrelation time for the plaquette and the gradient flow observables $t_0/a^2$ 
as shown in figure \ref{fig:autoc} by using the $\Gamma$-method \cite{Wolff:2003sm}.
We found a pion mass dependence given by
\begin{equation}
 \tau_{int}(m_\pi) = A \frac{1}{m_\pi^{b}}
\end{equation}
with $b = 2.2(5)$ case of the plaquette while $b=2.0(7)$ in case 
of the gradient flow observable $t_0$.
A possible explanation for the quark mass dependence of the autocorrelation 
time $\tau_{int}$ is a phase transition in case of finite twisted mass term for 
vanishing neutral pion masses. Although the isospin splitting is suppressed in 
our case, observables like the gradient flow observables 
shows an increase with inverse of the squared pion mass. This behavior is also seen in the PCAC mass,
where moderate integrated autocorrelation times
were found which can be clearly seen in the Monte Carlo history (see right panel of fig.~\ref{fig:tuning}).
However in other quantities like the pseudoscalar mass, the pseudoscalar decay constant
or nucleon observables $\tau_{int}$ is very small and a quark mass dependence can be not observed.

\bibliography{common.bib}

\end{document}